\documentclass[reprint,amsfonts,amssymb, amsmath,   nofootinbib,pra,aps,superscriptaddress, twocolumn,longbibliography]{revtex4-2}
\usepackage{titletoc}
\usepackage{tikz}
\usetikzlibrary{positioning}
\usepackage[colorlinks=true,citecolor=blue,linkcolor=magenta,linktocpage=true]{hyperref}
\usepackage{mathtools}
\usepackage{braket}
\usepackage{comment}
\usepackage{algorithm,algorithmicx,algpseudocode}
\usepackage{amsthm}
\usepackage{bbm}

\makeatletter
\newcommand{\doublewidetilde}[1]{{%
  \mathpalette\double@widetilde{#1}%
}}
\newcommand{\double@widetilde}[2]{%
  \sbox\z@{$\m@th#1\widetilde{#2}$}%
  \ht\z@=.9\ht\z@
  \widetilde{\box\z@}%
}
\makeatother
\makeatletter
\newsavebox{\@brx}
\newcommand{\llangle}[1][]{\savebox{\@brx}{\(\m@th{#1\langle}\)}%
  \mathopen{\copy\@brx\kern-0.5\wd\@brx\usebox{\@brx}}}
\newcommand{\rrangle}[1][]{\savebox{\@brx}{\(\m@th{#1\rangle}\)}%
  \mathclose{\copy\@brx\kern-0.5\wd\@brx\usebox{\@brx}}}
\makeatother

\newcommand{\spann}{{\rm Span}}

\newcommand{\trace}{{\rm Tr}}
\newcommand{\thv}{\vec{\theta}}

\newcommand\spn{\text{span}}
\def\tn{^{\otimes n}}
\def\ad{^{\dagger}}

\newcommand{\var}{{\rm Var}}
\newcommand{\prob}{p}
\DeclareMathOperator*{\E}{\mathbb{E}}

\newcommand{\mc}[1]{\mathcal{#1}}
\newcommand{\mbb}[1]{\mathbb{#1}}

\newcommand{\HH}{\mathcal{H}}
\newcommand{\BB}{\mathcal{B}}

\newcommand{\LL}{\mathcal{L}}

\newcommand{\MM}{\mathcal{M}}
\newcommand{\TT}{\mathcal{T}}
\newcommand{\AAA}{\POVM}
\newcommand{\VV}{\VDIST}

\def\HC{\mathcal{H}}

\newcommand{\lm}{\lambda }
\newcommand{\Tr}{{\rm Tr}}

\newcommand{\pihw}{\Pi_{h}}
\newcommand{\piw}{\Pi_{w}}

\newcommand{\pii}{ {\rm PI} }

\newcommand{\piopspace}{\BB^\pii}

\newcommand{\piop}[1][k]{B_{ \vec{#1} }}

\newcommand{\hw}{ h }

\usepackage{mathrsfs}
\def\EXP{\E_{V,w}}

\def\muv{\mu(V)}
\def\DIST{p_\rho(V, w)}
\def\VDIST{\mathscr{V}}
\def\POVM{\mathscr{P}}
\def\VIS{_{\textsf{vis}}}
\def\L{\mc{L}(\mc{H})}
\def\L{\mc{L}}

\def\LVIS{ \L\VIS }
\def\LG{\L^G}

\def\LSN{\mc{L}^{S_n}}
\def\LPI{\mathcal{L}^{\PI}}
\def\LINVIS{\L\VIS^{\perp}}
\def\HERM{\mc{B}(\HC)}
\def\HERM{\mc{B}}

\def\HERMG{\mc{B}_G(\HC)}
\def\HERMG{\mc{B}^G}

\def\var{{\rm Var}}

\def\U{\textsf{U}}
\def\SU{\textsf{SU}}

\def\PI{{\rm PI}}
\def\W{C}
\def\TC{\mathcal{T}}
\renewcommand\set[1]{\{ #1 \}}

\begin{document}
\title{Classical shadows with symmetries}
\author{Fr\'{e}d\'{e}ric Sauvage}
\affiliation{Quantinuum, Partnership House, Carlisle Place, London SW1P 1BX, United Kingdom}
\affiliation{Theoretical Division, Los Alamos National Laboratory, Los Alamos, New Mexico 87545, USA}
\author{Mart\'{i}n Larocca}
\affiliation{Theoretical Division, Los Alamos National Laboratory, Los Alamos, New Mexico 87545, USA}

\begin{abstract}
Classical shadows (CS) have emerged as a powerful way to estimate many properties of quantum states based on random measurements and classical post-processing. In their original formulation, they come with optimal (or close to) sampling complexity guarantees for generic states and generic observables. 
Still, it is natural to expect to even further lower sampling requirements when equipped with a priori knowledge regarding either the underlying state or the observables.
Here, we consider the case where such knowledge is provided in terms of symmetries of the unknown state \emph{or} of the observables. 
Criterion and guidelines for symmetric shadows are provided.
As a concrete example we focus on the case of permutation invariance (PI), and detail constructions of several families of PI-CSs. In particular, building on results obtained in the field of PI \emph{quantum tomography}, we develop and study shallow PI-CS protocol. Benefits of these symmetric CS are demonstrated compared to established CS protocols showcasing vastly improved performances.
\end{abstract}
\maketitle

\section{Introduction}
As the size of the quantum systems routinely manipulated has steadily increased over the years, full-fledged quantum tomography~\cite{vogel1989determination,paris2004quantum} has become impracticable;  more succinct ways to characterize quantum states, even if only partially, have been sought~\cite{gross2010quantum,cramer2010efficient,flammia2011direct,da2011practical,lanyon2017efficient,torlai2018neural,ahn2019adaptive,huang2020predicting,elben2022randomized}. 
Among them, classical shadows (CSs)~\cite{huang2020predicting} offer an efficient way to resolve accurately many properties of an unknown quantum state with only partial characterization.

As a key feature, CSs yield appealing scalings with respect to the number of properties to be estimated~\cite{huang2020predicting}: To achieve, with high probability, an additive error $\varepsilon$ in estimating the expectation values of a set of observables $\set{O_m}_{m=1}^M$, one needs a number of samples 
\begin{equation}\label{eq:sampling_scaling}
    S \in \mathcal{O}\Big( \frac{ \log(M)}{\varepsilon^2} \max_m \var [\hat{o}_m] \Big)\,,
\end{equation} 
where $\hat{o}_m$ is the estimate for $o_m = \Tr[\rho O_m]$.
In addition to the logarithmic and quadratic dependencies on $M$ and $\varepsilon^{-1}$ respectively, the required number of samples scale linearly in the individual \emph{single-shot variances} $\var [\hat{o}_m]_{\rm CS}$ of the estimates. Crucially, these variances depend on the specifics of the observables (e.g., their locality or norm) and on the choice of CS protocol. 
An appropriate choice of CS is thus one that reduces such variances for the situation of interest. 

Recent studies have shed lights on how different ensembles can result in different variances~\cite{zhao2021fermionic,low2022classical,wan2022matchgate,bertoni2022shallow,van2022hardware,hearth2023efficient,ippoliti2023classical}. 
A recurring theme in all these works is that one can reduce variances over some families of observables, but, at the cost of increasing them on others, and sometimes even losing the 
ability to make predictions for certain observables. This is to be expected, as efficient estimation of any arbitrary observable would enable full reconstruction (tomography) of the underlying state, that generically has to require exponentially many samples~\cite{flammia2011direct,brandao2020fast}.
Still, this shows that given some a priori knowledge on properties of the observables of interest (deviating from the generic situation) one can, in principle, design more efficient CSs.

The use of symmetries has permeated many aspects of sciences including tasks of quantum information and learning~\cite{gross1996role,zanardi2000stabilizing,chiu2016classification,walter2018symmetry,bronstein2021geometric,larocca2022group,meyer2023exploiting}. Generally speaking, symmetries often offer a mathematical framework to specify knowledge of a problem, and, ultimately, to simplify it. 
In the context of CSs, or more generally characterization of quantum systems, symmetries are a broad and natural language to specify families of observable of interest. It is thus natural to ask how symmetries can be used to tailor CSs. 

\subsubsection{Scope and contributions}
In this work, we consider the problem of estimating many properties of an unknown $n$-qubit state given the guarantee that the observables to be estimated obey some symmetries. 
As we shall see, this is equivalent to the case where the observables themselves are not symmetric but rather the unknown state is.
In this scenario, we establish a simple and general criterion for symmetric CS, that essentially requires that only \textit{symmetric} but \textit{sufficiently diverse} measurements are performed.

Symmetries come in different flavours such that implementations of symmetric CSs will inevitably vary from one (representation of a) group of symmetry to the other. Even for the same symmetries, various families of symmetric CSs can be designed.
Aiming toward concrete realizations, we detail two different approaches, one explicit and one implicit, to construct symmetric CS.

To exemplify these constructions, we then focus on a group of symmetry particularly well understood and of practical importance~\cite{harrow2005applications,harrow2013church,plesch2010efficient,alicki1988symmetry,keyl2001estimating,keyl2006quantum,haah2016sample,o2016efficient,o2017efficient,lidar2003decoherence,ouyang2014permutation,ouyang2016permutation,lipkin1965validity,yadin2023thermodynamics,markham2011entanglement,toth2012multipartite,davis2016approaching,linnemann2016quantum,kaubruegger2019variational,volkoff2022asymptotic,mills2019quantum,verdon2019quantumgraph,skolik2022equivariant,schatzki2022theoretical,mernyei2022equivariant,albrecht2023quantum,umeano2024geometric,dalyac2024graph}: the symmetric group
consisting in permutations of the elementary parts composing the underlying state.
In particular, building on results in the field of permutation invariant (PI) quantum tomography~\cite{toth2010permutationally,moroder2012permutationally}, we propose novel 
\textit{permutation-invariant classical shadows} (PI-CSs). As we will see, together with a decrease in the required sample complexity, such PI-CS schemes proposed have very low hardware demands: A single layer of correlated single qubit rotations followed by computational basis measurements. 
Other PI-CSs following Ref.~\cite{anschuetz2022efficient} are also explored, albeit requiring larger experimental overhead.

The PI-CS proposals are completed by characterization of the measurement channel realized (that is needed for the step of post-processing) and of the resulting variances.
Then, in numerical simulations, we demonstrate the superiority of PI-CS compared to established CSs. 
Notably, we observe a significant improvement in the sampling complexity with respect to the locality of the observables: For some operators acting non-trivially on the whole system (i.e., global operators) the sampling requirement of PI-CS is found to increase only linearly with the system size.
This is to be contrasted with the exponential scaling of more standard CSs. 
In other cases, with observables of bounded locality, PI-CS have even a sampling cost that \textit{decreases} with system size.

Of course, all these appealing features come at the cost of limiting ourselves to the estimation of a restricted set of observables (or, again, to arbitrary observables but restricted families of states). This is precisely the goal of symmetric CSs. 
Finally, we explore extensions of PI-CS to qudit systems and show that the present work can readily be extended to this setting.
Taken together, we expect these results to pave the way to the development of many more families of symmetric CSs.

\subsubsection{Related works}
One of the PI-CS presented builds on the field of PI quantum tomography~\cite{toth2010permutationally,moroder2012permutationally}. The main difference between quantum tomography and CSs (be them symmetric or not)
arises from the fact that the former aims at estimating all relevant entries of the underlying quantum state in a deterministic fashion, while the later only requires partial and random characterization. 
In particular, CSs remain valid (yield unbiased estimates) even in regimes where the number of measurements is smaller than the dimension of the underlying operator space under consideration. 
In turn, and as detailed in the result section, for most practical concerns PI-CSs yield improved sampling complexity.
Furthermore, quantum tomography relies on a fixed measurement basis.
In some cases, this choice can yield vastly different estimate accuracy, and needs to be optimized beforehand. 
Finally, and particularly relevant to hardware implementation, the randomization aspect inherent to CSs can facilitate mitigation of experimental errors~\cite{chen2021robust,koh2022classical,zhao2023group}.

A second PI-CS presented here follows the work of Ref.~\cite{anschuetz2019near}, and is  thoroughly exposed and assessed in numerical simulations. We found it similar in sampling complexity to the other PI-CSs, but to require substantially more circuit overhead. 
Also closely related to one of the  PI-CS presented here is the CS protocol of Ref.~\cite{van2022hardware}. 
As we shall see, the difference in both our and their protocols lies on the measurements performed and post-processing. 
In particular, the measurements adopted here is key in our CSs being symmetric and introduce intentional coarse graining of the measurement data.
Pursuing comparison with Ref.~\cite{van2022hardware}, we showcase scenarios where symmetry knowledge can be incorporated \emph{a posteriori} and still improve the estimation accuracy: With an already acquired random-measurement dataset one can enforce symmetry at the classical post-processing stage and improve accuracy of the estimates.

Finally we note that a few existing works could be cast under the umbrella of symmetric CSs.
For instance, the protocol of Ref.~\cite{hearth2023efficient} qualifies as symmetric CS, and as much as Ref.~\cite{van2022hardware} is motivated by hardware restrictions. In the present work we rather impose such restrictions on purpose.
On the other hand Ref~\cite{wan2022matchgate, zhao2021fermionic} may also be cast as symmetric CS, but, with symmetries manifesting in different way than studied here (namely, quadratic symmetries rather than linear ones~\cite{zimboras2015symmetry}). This is discussed further later on. 

\section{Symmetry and Classical shadows}
We start by introducing notation (Sec.~\ref{sec:notations}) and reviewing the CS formalism (Sec.~\ref{sec:cs}), including details of several existing CSs. Then, relevant elements for the treatment of symmetries are presented (Sec.~\ref{sec:symm}).
This leads us to define the simple and broadly applicable concept of symmetric CSs, and to provide guidelines toward their constructions (Sec.~\ref{sec:symm_cs}). Further background and justifications are provided in App.~\ref{app:background} of the appendices.

\subsection{Notations}\label{sec:notations}
Given $\HH$ a Hilbert space of dimension $d$, let $\mathcal{L}(\HH)$ be the space of linear operators on $\HH$ which includes, as subspaces, the Hermitian operators $\mathcal{B}(\mathcal{H}):= \set{B \in \mathcal{L}(\HC) \,|\, B = B^{\dagger}}$ and the group of unitary operators $\textsf{U}(\HC) := \set{U \in \mathcal{L}(\HC) \,|\, U U^{\dagger} = U^{\dagger} U = I_{\HC}}$, with $I_{\HC}$ the identity on $\mathcal{H}$. The group of special unitaries are those unitaries with unit determinant and denoted $\textsf{SU}(\HH)$. 
Sometimes, we identify unitaries through the dimension on the space they act on with, e.g., $\textsf{SU}(2)$ for the group of special unitaries acting on a single qubit.
Unless otherwise stated we will focus on $n$-qubit systems with $\HH:=(\mathbb{C}^2)\tn$ and $d\coloneqq \dim(\HC)=2^n$. In such case we drop the $\HH$ in our notations. For instance, $\mathcal{L}$ is the space of operators acting on $n$-qubit states.
Operator spaces are endowed with the Hilbert-Schmidt inner product $\langle A, B\rangle := \trace[A^{\dag} B]$ that induces the (Froebenius) norm $||A||^2:=\trace[A^{\dag} A]$. We also make use of the spectral norm denoted $||A||_{\infty}$.

\subsection{Classical shadows}\label{sec:cs}
CSs aim at estimating properties of a unknown $n$-qubit state $\rho$. By properties we mean the expectation values $o_m := \trace[\rho O_m]$ for a set of observables $\set{O_m}_{m\in [M]}$ with each $O_m \in \BB$ and where $[M]$ denotes the set $\set{1, \ldots, M}$. 

A CS protocol~\cite{huang2020predicting} consists in two basic steps:
\begin{enumerate}
    \item a data acquisition stage where one collects data obtained from a set of \textit{random} measurements of $\rho$. 
    \item a classical post-processing phase where the measurement data are used to evaluate estimates $\hat{o}_m$ of the expectation values $o_m$.
\end{enumerate}
Notably, these steps can be fully decoupled such that the specification of the observables to be resolved may only become available after having measured the system.

\subsubsection{Data acquisition}\label{sec:cs_data}
During data acquisition, a total of $S$ random measurements are performed. Each random measurement consists in (i) the application of a \emph{random} unitary $V$ to $\rho$, with $V$ drawn from an ensemble $\VDIST=(H,\mu)$ where $H\subset \textsf{SU}(d)$ is a subgroup of unitaries, and $\mu$ a probability measure $\mu:H \xrightarrow[]{} [0,1]$\footnote{Throughout this work, $\mu$ will be the uniform distribution (Haar distribution on Lie group).
Having fixed a distribution, we use $\VDIST$ both to refer to the group of unitaries $H$ and to the ensemble.
}, followed by (ii) a \emph{fixed} measurement specified by 
a set of projectors\footnote{We limit ourselves to projectors, but such measurement can be generalized to positive-operator value measure (POVM).} $\POVM = \{ \Pi_w\}_{w \in W}$ such that $\sum_w \Pi_w = I$.
Hence, the outcome pair $(V, w)$ yield by one of such 
random measurement
follows the joint distribution 
\begin{align}\label{eq:proba_w_V}
\begin{split}
      \DIST = \mu(V) \trace[V^{\dagger} \rho V \piw ]\,.
\end{split}
\end{align}
We denote as $\EXP$ expectation values taken over such distribution. The set of $S$ outcomes obtained are recorded in a dataset $\mathcal{D}=\set{(w_s, V_s)}_{s \in [S]}$. 

The choice of $\VDIST$ along with the measurement $\POVM$ determine a CS protocol. The \emph{measurement channel} associated with a given CS protocol is defined as
\begin{align}\label{eq:mc}
\begin{split}
     \MM( \rho ) &:= \EXP \bigg[V^{\dagger} \piw V \bigg] \\
     &= \sum_w \int_{V}  \muv \trace[ V\rho V^{\dagger} \piw] V^{\dagger} \piw V.
\end{split}
\end{align}
This definition is readily extended to any input operator $X$, rather than only quantum states $\rho$. It can be verified From Eq.~\eqref{eq:mc} that $\MM$ is self-conjugate such that $\trace[X^{\dagger} \MM(Y)]=\trace[\MM(X)^{\dagger} Y]$ for any $X$ and $Y \in \mathcal{L}$.

\subsubsection{Post-processing and statistical properties}\label{sec:cs_post}
Consider an observable $O$, an outcome $(V, w)\in \mathcal{D}$, and define as $\hat{\rho}:= \MM^{-1}(V^{\dagger} \Pi_w V)$ the single-shot CS estimate of $\rho$ corresponding to such outcome.
Note that we are assuming $\MM$ is invertible (we will come back to this assumption soon).
By construction, $\hat{\rho}$ is unbiased as
\begin{align}
    \begin{split}
    \EXP[\hat{\rho}] &= \EXP \bigg[\MM^{-1}\Big(V^{\dagger} \piw V  \Big) \bigg] = \MM^{-1}(\MM(\rho)) 
    \\&=\rho.
    \end{split}
\end{align}
It follows by linearity that the single-shot CS estimate
\begin{equation}\label{eq:cs_obs}
    \hat{o}:= \trace [O \hat{\rho}] = \trace [O \MM^{-1}(V^{\dagger}\Pi_w V)]
\end{equation} 
is unbiased toward $o:= \trace [O \rho]$,
and has variance
\begin{align}\label{eq:var}
\begin{split}
     \var[\hat{o}]_{\rm CS} &:= \EXP[\hat{o}^2] - \Big(\EXP[\hat{o}] \Big)^2 \\&= \EXP \Big[ \trace \Big[ \mathcal{M}^{-1}\Big(V^{\dagger} \Pi_w V\Big) O \Big]^2 \Big] - o^2.
\end{split}
\end{align}

The previous discussion for a single observable $O$ is readily extended to any set of observables $\set{O_m}_{m\in [M]}$ and we denote as $\hat{o}_m$ the corresponding estimates.
Furthermore, using the entire dataset $\mathcal{D}$ of $S$ measurements, rather than a single one, yields more precise estimates denoted $\hat{o}_m (S)$. 
While empirical averages could be used, other statistical methods exist. 
In particular \emph{Median of Means} are often favoured in the context of CS as they allow to construct estimates that, with high probability, ensure that $ |\hat{o}_m(S) - o_m|<\varepsilon$ for all $m$, given the number of samples reported in Eq.~\eqref{eq:sampling_scaling}.

\subsubsection{Compatible observables}
From the previous discussion, it would seem that any observable can be estimated through CSs. This however necessarily depends on the random measurements performed.
For instance, consider the trivial unitary ensemble $\VDIST=\{ I\}$, and computational basis measurements
\begin{equation}\label{eq:cb_meas}
    \POVM_{\rm cb} :=\{ \ket{w}\bra{w} \}_{w \in \set{0,1}^n} .
\end{equation}
It should be clear that outcomes of these measurements can only be used to  estimate (without a bias) observables that are diagonal in the computational basis. 

More generally, given a unitary ensemble $\VDIST$ and measurement $\POVM$ 
the outcomes $(w,V)$ can only be employed to estimate expectation values of linear combinations of $V^{\dagger} \Pi_w V$. Hence, the \emph{visible operator space} of a CS protocol is, as defined in Ref.~\cite{van2022hardware}, the operator subspace
\begin{equation}\label{eq:vis}
\LVIS = \spann \set{V^{\dagger} \Pi_w V\,|\, V \in \VDIST \text{ and } \Pi_w \in \POVM} \subseteq \L
\end{equation}
consisting of all observables that can be estimated by the CS protocol. We denote as $\LINVIS$ its orthogonal complement such that $\LVIS \oplus\LINVIS = \L$.

Relating this visible space to the measurement channel, one can see from  Eq.~\eqref{eq:mc} that the channel output is a linear combination of $V^{\dagger}\Pi_w V$, ${\rm Im}(\MM) = \LVIS$, while any input component in $\LINVIS$ does not contribute, ${\rm Ker}(\MM) =\LINVIS$.
That is, given a orthonormal basis $\set{B_k}$ for $\LVIS$ and Hermitian $X\in\HERM$ the channel has the form
\begin{equation}\label{eq:mat_meas_channel}
    \MM(X) = \sum_{k,k'} C_{k,k'} \trace[B_k^{\dagger} X] B_{k'},
\end{equation}
fully specified by a matrix $C$
of real-valued coefficients $C_{k,k'}\in \mathbb{R}$.

\subsubsection{Tomographic completeness and some known CSs}\label{subsec-clifford}
\emph{Tomographic-complete} CSs are those CSs where $\LVIS = \L$. These include, as prominent examples, \textit{local Clifford} (LC) and \textit{global Clifford} (GC) classical shadows~\cite{huang2020predicting}. Denoting $\textsf{Cl}(2^k)$ the Clifford group on $k$-qubits, these two protocols correspond to the unitary ensembles
\begin{align}\label{eq:RP_RC}
\begin{split} 
    &\VDIST_{{\rm LC} } := \{ V=W_1 \otimes \hdots \otimes W_n \,|\, W_i \in \textsf{Cl}(2)\} \textrm{ and  }  \\ 
     &\VDIST_{{\rm GC} } := \{V \,|\, V \in \textsf{Cl}(2^n) \},
\end{split}
\end{align}
respectively, followed by computational basis measurements as per Eq.~\eqref{eq:cb_meas}. Variances of the CS estimates are bounded as 
\begin{align}\label{eq:var_RP_RC}
\begin{split}    
    \var[\hat{o}]_{\rm LC} &\leq 4^{{\rm loc}(O)} || O ||^2_{\infty} \textrm{ and  } \\
    \var[\hat{o}]_{\rm GC} &\leq 3 ||O||^2,\\
\end{split}
\end{align}
respectively, where ${\rm loc}(O)$ is the locality of the observable $O$, i.e., the number of qubits on which it acts non trivially.

Recent works have studied \emph{non-tomographic-complete} CSs~\cite{van2022hardware, hearth2023efficient,ippoliti2023classical} where $\LVIS \subsetneq \L$. In such case $\MM^{-1}$ is the \emph{pseudo-inverse} of the superoperator channel $\MM$, such that their composition $\mathcal{M} \circ \mathcal{M}^{-1}$ is the orthogonal projector onto $\LVIS$. 

Reduction of the visible space arises from specific choice of the unitary ensemble $\VDIST$ and also from the choice of the measurement $\POVM$. 
Motivated by hardware limitations, Ref.~\cite{van2022hardware} studies CSs with global control, whereby one cannot control individually parts of the system, as can happen in ultra-cold atoms setups. For instance, it considers the ensemble of correlated local unitaries (CLU)
\begin{align}\label{eq:collective_V}
\begin{split}
     &\VDIST_{{\rm CLU}}=\set{ W^{\otimes n} \,|\, W \in \textsf{SU}(2)}
\end{split}
\end{align}
together with measurements in the computational basis. Note that, while the qubits are controlled globally (correlated), it is assumed that they can be measured individually. This yields a visible space $\LVIS$ with $\dim(\LVIS)=2^n(2n^3 + 7n^2+8)/8$. This roughly corresponds to a square-root reduction as compared to the space $\L$, with $\dim(\L)=4^n$, visible to tomographic-complete CSs.

In the same vein, Ref.~\cite{hearth2023efficient} restrict themselves to number-preserving unitaries as available with hardcore bosons experiments yielding again a reduced visible space. While these works were aimed at understanding CSs in the presence of hardware restrictions, our goal will rather be to intentionally reduce (i.e., tailor) the visible space to the space of relevant symmetric operators. But first, we need to formalize what we mean by symmetries.

\subsection{Symmetries}\label{sec:symm}
Let $G$ be a group of \textit{symmetries} and $U:G\xrightarrow{} \U(\HC)$ a \textit{unitary representation} for $G$ on $\HC$. The representation $U$ specifies how the group elements $g \in G$ act on quantum states $\ket{\psi} \in \mathcal{H}$, through $g \cdot \ket{\psi} = U(g) \ket{\psi}$, while preserving the group structure: $U(g)U(g')=U(g g')$ for any $g, g' \in G$. 
The action of the symmetry group $G$ on pure states is readily extended to an action on linear operators $A \in \mathcal{L}$ through conjugation $g \cdot A = U(g) A U^{\dagger}(g)$. In the following, we use the shorthand notation $U_g := U(g)$. 

An operator $A \in \L$ is said to be \textit{symmetric}, or \textit{invariant}, (under $G$) whenever $G$ acts trivially on it: $g\cdot A = U_g A U_g\ad=A$ for all $g \in G$. This condition is equivalent to $[U_g,A]=0$ for all $g$. 
The subspace of symmetric operators is defined as
\begin{equation}\label{eq:linear_symm}
    \mathcal{L}^G := \left\{ A \in \L \, | \,  [A, U_g]=0,\;  \forall g \in G 
    \right\}.    
\end{equation}
In analogy, we denote as $\HERMG = \HERM \cap \LG \subset \LG$ the subspace of \textit{Hermitian} symmetric operators.
Two concepts, representation theory and twirl, are particularly useful when dealing with symmetric operators. In particular, they can be leveraged to identify concrete bases for $\mathcal{L}^{G}$ and, as used later on, to design symmetric CSs. 
 
\subsubsection{Representation theory}
For compact groups of symmetries, the invertible linear operators corresponding to represented group elements $U(g)$, for the different $g\in G$, can be simultaneously block-diagonalized into fundamental building blocks called 
\textit{irreducible representations} (or \emph{irreps}). Denoting these irreps as $r_{\lambda}(g) \in \textsf{U}(\mathbb{C}^{d_{\lambda}})$, with $d_\lm$ the irrep dimension, the aforementioned block-decomposition take the form
\begin{equation}\label{eq:irrep_decomp}
    U_g \cong \bigoplus_{\lambda} r_{\lambda}(g) \otimes I_{m_{\lambda}},
\end{equation}
with $m_\lm$ the multiplicity of the irrep $\lm$ and $I_{m_{\lambda}}$ the identity on $\mathbb{C}^{m_{\lambda}}$. Eq.~\eqref{eq:irrep_decomp} indicates that, for an appropriate change of basis ($\cong$), representations of \emph{any} $g$ adopt a similar \emph{block diagonal} form where each of the distinct blocks $r_{\lambda}(g)$ is repeated $m_{\lambda}$ times.
Under the same change of basis, the Hilbert space decomposes as
\begin{equation}\label{eq:irrep_decomp_H}
    \mathcal{H} \cong \bigoplus_{\lambda} \mathcal{H}_{\lambda} \otimes \mathbb{C}^{m_{\lambda}} 
\end{equation}
and we label elements of the new basis (called here the \emph{irrep} basis) as $\ket{\lambda, i_{\lm}, j_{\lm}}$ with $i_{\lambda} \in [d_{\lambda}]$ labeling a basis of $H_{\lambda}$ and $j_{\lm}\in[m_{\lambda}]$ indexing its copies. 

From Eq.~\eqref{eq:irrep_decomp}, one can see that matrices $A$ of the form 
\begin{equation}\label{eq:irrep_decomp_comm}
    A \cong \bigoplus_{\lambda} I_{d_{\lambda}} \otimes A_{\lambda}, 
\end{equation}
for $I_{d_{\lambda}}$ the identity on $\mathbb{C}^{d_{\lambda}}$ and arbitrary $A_\lambda \in \mathcal{L}(\mathbb{C}^{m_{\lambda}}$) commute with all the $U_g$. That is, $A$ are \textit{symmetric} 
operators. In fact, by virtue of the Schur's Lemma~\cite{fulton1991representation}, one can show that any operator $A \in \mathcal{L}^{G}$ has to be of the form Eq.~\eqref{eq:irrep_decomp_comm}. Hence, with 
\begin{equation}\label{eq:basis_rep_th}
     B_{\lm,j_\lm,j_\lm'} := \sum_{i_{\lm}=1}^{d_\lm} \ket{\lambda, i_{\lm}, j_{\lm}}\bra{\lambda, i_{\lm}, j'_{\lm}} 
\end{equation}
the set $ \Bigr\{ B_{\lm,j_\lm,j_\lm'} \Bigr\}_{\lm, j_{\lm}, j'_{\lm}=1}^{m_\lm} $ forms an orthogonal basis of the symmetric subspace $\LG$, that we call \emph{irrep} operator basis, and we obtain ${\rm dim}(\LG)= \sum_{\lambda} m^2_{\lambda}$.

\subsubsection{Twirl}
Non-symmetric objects can be made symmetric by averaging them under the  group action. For operators, such a \emph{twirl}, $\mathcal{T}_G:\LL \mapsto \LL^G$, is defined as 
\begin{equation}\label{eq:def_twirl}
    \mathcal{T}_G(A) := \frac{1}{|G|} \sum_{g \in G} g \cdot A = \frac{1}{|G|} \sum_{g \in G} U_g A U^{\dagger}_g
\end{equation}
We write $A^G := \mathcal{T}_{G}(A)$.
Notably, Eq.~\eqref{eq:def_twirl} is an orthogonal projection onto $\LG$
such that 
$A^G\in \LG$, and $(A^G)^G=A^G$ for any $A \in \mathcal{L}$. 
It follows that a basis of $\LG$ can be obtained by twirling any basis of $\L$.
However, as opposed to Eq.~\eqref{eq:basis_rep_th}, the resulting basis may not be orthogonal and can be overcomplete. Finally, given properties of the twirl and linearity of the trace, one can show that for any state $\rho$ and observable $O$,
\begin{equation}\label{eq:equiv_expectations}
    \trace[O^G \rho] = \trace[O \rho^G] = \trace[O^G\rho^G].
\end{equation}

\subsection{Symmetric Classical Shadows}\label{sec:symm_cs}
Crucially, Eq.~\eqref{eq:equiv_expectations} shows that to compute properties of a state given the guarantee that either (i) the observable is symmetric or that (ii) the underlying state is symmetric or (iii) both, we only need to characterize the symmetric part $\rho^G$ of the state $\rho$; the non-symmetric part is completely un-informative.
This motivates the general and simple criterion for \emph{symmetric CS}:
\begin{equation}\label{eq:symm_CS_rule}
    \LVIS = \LG
\end{equation}
that requires that the visible space in Eq.~\eqref{eq:vis} corresponds exactly to the symmetric operator space of Eq.~\eqref{eq:linear_symm}.

\begin{figure}[t]
\centering
\includegraphics[width=\columnwidth]{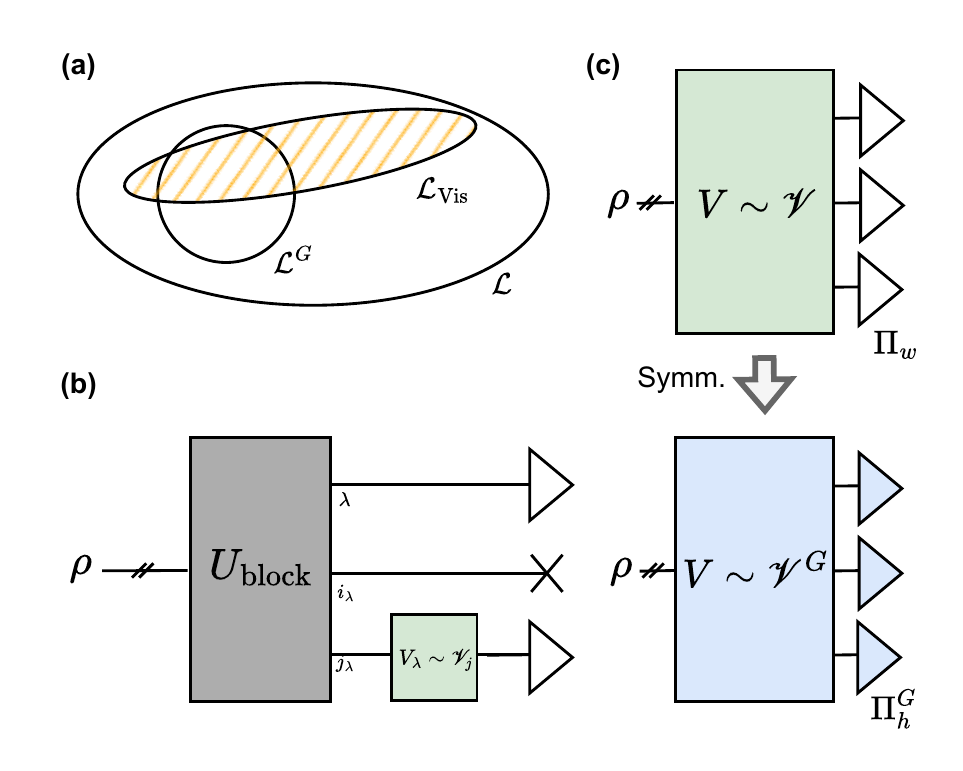}
\caption{
\textbf{Symmetric CSs:} \textbf{(a)} The aim of symmetric CSs is to align the visible space of a CS with the space of symmetric observables, i.e., to satisfy Eq.~\eqref{eq:symm_CS_rule}. This can be achieved \textbf{(b)} by explicitly performing block-diagonalization of the action of the symmetries and additional partial random measurements or by \textbf{(c)} symmetrizing non-symmetric CS protocols.}
\label{fig:symm_cs}
\end{figure}

Instantiation of concrete symmetric CSs necessitates designing measurements $\POVM$ and unitary ensemble $\VDIST$ such that Eq.~\eqref{eq:symm_CS_rule} is satisfied. We stress that different CS protocols can satisfy this condition, such that families of symmetric CS exist (as much as there exists many CS protocols that are tomographic-complete). We now detail two distinct approaches to design symmetric CSs that are sketched in Fig.~\ref{fig:symm_cs}.

\subsubsection{Symmetric CS through block-diagonalization}\label{sec:cs_symm_rep} 
From the representation theory perspective a very explicit design principle emerges as we aim at measuring observables of the form Eq.~\eqref{eq:basis_rep_th}, or linear combinations of such. Such protocol was first sketched in Ref.\cite{anschuetz2022efficient} and is detailed further in the following.

Let us assume that we have access to the transform $U_{\rm block}$ that performs the change of basis to the irrep basis, in which symmetry representations and symmetric operators adopt the block-diagonal forms of  Eq.~\eqref{eq:irrep_decomp} and Eq.~\eqref{eq:irrep_decomp_comm} respectively. 
Furthermore, assume that such transform encodes the representation-theory labels in separable registers such that $\ket{\lambda, i_\lambda, j_{\lambda}} = \ket{\lambda} \otimes \ket{i_\lambda} \otimes \ket{j_{\lambda}}$. 
We refer to the registers containing the states $\ket{\lm}$, $\ket{i_\lm}$ and $\ket{j_\lm}$ as the \textit{irrep}, the \textit{dimension} and the \textit{multiplicity} registers respectively.
Furthermore let us define $m:=2^{\lfloor \log_2( \max_{\lm} m_\lm )\rfloor}\approx \max_\lm m_\lm$ the dimension of the multiplicity register such that it is composed of $k=\log_2(m)$ qubits\footnote{Given that the number of labels $i_\lm$ and $j_\lm$ depends on $\lm$, these registers need to accommodate for the largest values of $m_\lm$ and $d_\lm$ over all $\lm$. In circuit implementation, the transform will require introducing additional ancilla qubits.}.

A first family of symmetric CSs, dubbed \emph{block-CS}, is illustrated in Fig.\ref{fig:symm_cs}(b) and constructed as follows: 
(i) apply $U_{\rm block}$,
(ii) measure the irrep register, obtaining an outcome $\lm$,
(iii) apply a random unitary $V_\lm \in H \subseteq \SU(m_\lm)$ 
on the multiplicity register and measure this register, obtaining an outcome $j_\lm$.
Notably, the dimension register is never measured as it encodes redundancy due to the symmetries.

In the irrep basis (after the transform), the random unitaries have the form $V=I \otimes I \otimes V_\lm$ while the measurements have the form $\Pi_{\lm, j_\lm} = \sum_{i_\lm} \ket{\lm}\bra{\lm} \otimes \ket{i_\lm}\bra{i_\lm} \otimes \ket{j_\lm}\bra{j_\lm}$, with the sum accounting for tracing out the dimension register. 
As such, the visible space is spanned by the operators
\begin{equation}\label{eq:vis_span_irrep_cs}
V^{\dagger} \Pi_{\lm, j_{\lm}} V = \sum_{i_\lm} \ket{\lambda} \bra{\lambda} \otimes \ket{i_{\lm}} \bra{i_{\lm}} \otimes V_\lm^{\dag} \ket{j_{\lm}} \bra{j_{\lm}} V_\lm, 
\end{equation}
that, by construction, all comply with Eq.~\eqref{eq:irrep_decomp_comm}. Hence, the visible space of the block-CS is a subset of the symmetric operator space. 
To satisfy the equality in Eq.~\eqref{eq:symm_CS_rule}, we simply need to ensure that the unitary ensemble together with the measurement performed  on the $k$ qubits of the multiplicity register are tomographic-complete in this register. 
In particular, we can choose the LC or GC shadows, with ensembles defined in Eq.~\eqref{eq:RP_RC},
and computational basis measurements. For these, we can readily derive bounds on the variances entailed as now detailed.

Given that only symmetric parts contribute, as per Eq.~\eqref{eq:equiv_expectations}, and to facilitate the exposition, we assume that both the unknown state $\rho$ and the observable $O$ are symmetric. According to Eq.~\eqref{eq:irrep_decomp_comm}, we can decompose them as $\rho \cong \bigoplus_{\lambda} I_{m_{\lambda}} \otimes \rho_{\lambda}$ and  $O \cong \bigoplus_{\lambda} I_{m_{\lambda}} \otimes O_{\lambda}$ such that 
\begin{equation}
o= \trace[O \rho]= \sum_{\lm} m_\lm \trace[O_\lm \rho_\lm].
\end{equation}
Consider a measurement outcome pair $(\lm, j_{\lm})$ from block-CS. 
First, note that an outcome $\lm$ happens with probability $\prob(\lm):=m_{\lm} \trace[\rho_\lm]$ and, after tracing out the dimension register, results in the (normalized) state $\tilde{\rho}_{\lm} := \rho_{\lm}/\trace[\rho_{\lm}]$.
Thus, through performing CS on the multiplicity register (with either the LC or GC ensemble), one gets an unbiased estimate $\hat{o}_\lm$ of $o_\lm = \trace[O_\lm \tilde{\rho}_{\lm}]$ with variance $\var[\hat{o}_\lm]_{\rm CS}$.

Denote as $\hat{o}$ the random variable taking value $\hat{o}_\lm$ given a randomly obtained $\lm$.
This estimate is unbiased, as in average $\E_\lm [\hat{o}] :=  \sum_{\lm} \prob(\lm) \E[\hat{o}_{\lm}] = o$, and has variance
\begin{align}\label{eq:var_block_cs_abstract}
\begin{split}    
    \var[\hat{o}]_{\rm block-CS} &= \left( \sum_{\lm} \prob(\lm) \var[\hat{o}_\lm]_{\rm CS}\right) + \var_{\lm}[o_\lm].
\end{split}
\end{align}
The first summand of the RHS in Eq.~\eqref{eq:var_block_cs_abstract} is bounded by $\max_{\lm} \var[\hat{o}_\lm]_{\rm CS}$. The second term in the RHS is the variance of a random variable taking value $o_{\lm}$ with probability $\prob(\lm)$. It is bounded by $\max_{\lm} |o_{\lm}|^2$ (that depends on $\tilde{\rho}_{\lm}$) and thus by $\max_{\lm} ||O_{\lm}||_{\infty}^2$ (for arbitrary $\tilde{\rho}_{\lm}$).  

Given that the operator $O_{\lm}$ acts on the multiplicity register of dimension $m$ and that $||O_{\lm}||_{\infty} \leq||O||_{\infty}$, we get  $||O_{\lm}||^2 \leq m^2||O_{\lm}||_{\infty}^2 \leq m^2||O||_{\infty}^2$. Finally recalling bounds on the CSs variances for the RC and RP ensembles of Eq.\eqref{eq:var_RP_RC}, we obtain the bounds 
\begin{align}\label{eq:var_block_CS}
\begin{split}    
    \var[\hat{o}]_{\rm block-RP} &\leq  (m^2+1)||O||^2_\infty \; \text{and }\\
    \var[\hat{o}]_{\rm block-RC} &\leq 3( m^2 + 1) ||O||^2_{\infty},
\end{split}
\end{align}
for any state $\rho$. These both scale as $m^2$, highlighting the role of the dimension $m \approx \max_{\lm} m_{\lm}$ in the sampling efficiency. Note that $m\leq d$, with equality occurring in the case when no symmetries are present, and $m<d$ for non-trivial symmetries.

On one hand, this approach towards the design of symmetric CS has the advantage of being explicit and readily ensures that Eq.~\eqref{eq:symm_CS_rule} is satisfied. Furthermore bounds on the variances can be readily derived, as given in Eq.~\eqref{eq:var_block_CS}.
On the other hand, it requires that a circuit implementation of the transform $U_{\rm block}$ is known in the first place, and even when known, such transform may introduce unnecessary circuit overhead. 
Hence, in the following we explore another approach.

\subsubsection{Symmetrized CS via twirling}\label{sec:symm_cs_symm}
As an alternative and sketched in Fig.~\ref{fig:symm_cs}(c), we can \emph{symmetrize} non-symmetric CS protocols. The aim here is to ensure that both the unitaries and the measurements are made symmetric; this can be achieved by means of the twirl defined in Eq.~\eqref{eq:def_twirl}. Similar strategies have been explored in the context of Geometric Quantum Machine Learning~\cite{meyer2022exploiting,nguyen2022atheory}.

Let us consider, as a reference, a non-symmetric CS protocol with unitary ensemble $\VDIST$ and measurement $\POVM$. We now detail how their symmetrized version, dubbed \emph{symm-CS} is obtained.
First, we can readily symmetrize the measurement by twirling each of the projectors involved. This yield the symmetrized measurement 
\begin{equation}
\POVM^G=\set{\TC_G(\Pi)}_{\Pi\in \POVM}.    
\end{equation}

We could envision the same approach for the unitary ensemble, however note that the twirl of an unitary will, in general, not be unitary. 
Non unitary operations could still be implemented through, for instance, linear combination of unitaries (LCU) techniques~\cite{childs2012hamiltonian}, but will require additional circuit overhead (including additional qubits, post-selection, and controlled operations).
Rather, we parameterize the group of unitaries through their generators $\mathcal{G}=\set{A_i\in \BB}$ such that  $\VDIST=\{V=\exp[-i \sum_i \theta_i A_i]\}$ for appropriate choice of parameters $\theta_i$.
Denoting the twirled generators $\mathcal{G}^{G}:=\{ \mathcal{T}^G(A_i)\}_{A_i \in \mathcal{G}}$, the symmetrized unitary ensemble is defined as 
\begin{equation}
    \mathcal{V}^{G}:=\set{\exp[-i \sum_i \theta_i A_i]}_{A_i \in \VDIST^G},
\end{equation}
with each element being both symmetric and unitary.

By construction, the unitary ensemble and the measurement obtained satisfy, $[V, U_g]=0$ and $[\Pi, U_g]=0$ for all $V\in \VDIST^G$, $\Pi \in \POVM^G$ and $g\in G$, guaranteeing that $\LVIS \subset \LG$.
However, the symmetrization does not ensure the equality required by our criterion for symmetric CS in Eq.~\eqref{eq:symm_CS_rule}. This has to be verified case by case and will depend on the proposal for $\VV$ and $\AAA$ in the first place. Furthermore, derivations of the resulting measurement channels and variances will also need to be assessed for each symmetrized protocol.
It is also the case that implementation of the symmetric measurement and unitaries could require extra effort compared to their non-symmetric alter-ego.
Overall, this shows that this symmetrization approach is more ad-hoc than the block-diagonalization of Sec.~\ref{sec:cs_symm_rep}.
Despite these complications, and as we shall soon see, it can yield symm-CSs much more compact than the block-CSs detailed earlier.

\section{Permutation-invariant Classical Shadows}\label{eq:pi_cs}
Having laid out our program of symmetric CSs, we now instantiate it for a particular group of symmetries: $S_n$, the symmetric group of permutations of $n$ elements. 
Such symmetries occur, under different guise, in many domains of quantum physics including 
quantum information~\cite{ harrow2005applications,harrow2013church,plesch2010efficient,alicki1988symmetry,keyl2001estimating,keyl2006quantum,haah2016sample,o2016efficient,o2017efficient}, error correction~\cite{lidar2003decoherence,ouyang2014permutation,ouyang2016permutation}, condensed-matter~\cite{lipkin1965validity,yadin2023thermodynamics}, theory of entanglement~\cite{ markham2011entanglement,toth2012multipartite} metrology~\cite{davis2016approaching,linnemann2016quantum,kaubruegger2019variational,volkoff2022asymptotic} and machine-learning~\cite{mills2019quantum,verdon2019quantumgraph,skolik2022equivariant,schatzki2022theoretical,mernyei2022equivariant,albrecht2023quantum,umeano2024geometric,dalyac2024graph}.

In Sec.~\ref{sec:pi} we recall generalities of the symmetric group. Further details are provided in App.~\ref{app:bck_sn}. Then, following the principles presented earlier we present two families of \emph{permutation-invariant CSs} (PI-CSs), block-CS that necessitates deep quantum circuits (at least linear depth in the system size) in Sec.~\ref{sec:pi_protocol_1} and symm-CS that only requires shallow quantum circuits (one layer of single-qubit rotations) in Sec.~\ref{sec:pi_protocol_2}. 
The former readily follows from Sec.~\ref{sec:cs_symm_rep} while the later, based on Sec.~\ref{sec:symm_cs_symm}, requires to additionally work out its explicit details.
In particular, we prove that one layer of single-qubit rotations is enough to satisfy Eq.~\eqref{eq:symm_CS_rule}. Similar arguments support PI quantum tomography~\cite{toth2010permutationally} but, to our knowledge, were not formally proven. 
Furthermore, for these shallow PI-CS, we derive in Apps.~\ref{app:mc_var_symm_pics} and~\ref{app:mc_schur_basis} the measurement channel required for post-processing and expressions for the variances entailed.
Finally, through comprehensive comparison we confirm advantages and scalability of the proposed PI-CS in the numerical demonstrations of Sec.~\ref{sec:pi_dem}.

\subsection{Permutation invariance}\label{sec:pi}
The symmetric group $S_n$ is naturally thought of as the set of all the permutations over a set $[n]:=\set{1, \hdots, n}$ of $n$ indices. Action of a permutation $\sigma \in S_n$ on any $n$-qubit separable state
\begin{equation}\label{eq:action_Sn}
    U_{\sigma} \big( \ket{i_1} \otimes \hdots \otimes \ket{i_n} \big) = \ket{i_{\sigma^{-1}(1)}} \otimes \hdots \otimes \ket{i_{\sigma^{-1}(n)}},
\end{equation}
and extended by linearity to arbitrary state $\ket{\psi} \in \HC= (\mbb{C}^2)\tn$.
Following the previous discussion, the action of $\sigma$ on an operator $A$ is given by $\sigma \cdot A:=U_{\sigma}AU_{\sigma}^{\dagger}$.
The space of PI operators corresponds to those symmetric under $G=S_n$, $\mathcal{L}^{\PI} := \LSN$,
and two bases for this space are now presented.
\subsubsection{Symmetrized Pauli string}\label{sec:pi_symm_paulis}
A basis for $\LPI$ can be obtained by twirling a basis for $\mathcal{L}$. Taking the $4^n$ Pauli strings $\set{X, Y, Z, I}^{\otimes n}$ as a basis of $\LL$, we can see that all Pauli strings with the same composition (i.e., same number $k_\alpha$ of Pauli operators $\alpha \in \set{X, Y, Z, I}$ independent of their orderings) map, under the twirl, to the same operator. Hence, we can label the resulting operators by a vector of non-negative integers $\vec{k} = (k_X, k_Y, k_Z, k_I)$ that are valid \emph{compositions} provided that  $|\vec{k}|:= \sum_{\alpha} k_\alpha = n$. 

Explicitly, we get that the \emph{symmetrized Pauli strings}
\begin{equation}\label{eq:A_k}
    B_{\vec{k}} = \TC_{S_n} (X^{\otimes k_X} Y^{\otimes k_Y} Z^{\otimes k_Z} I^{\otimes k_I})
\end{equation}
form a basis of $\LPI$. Given that each $B_{\vec{k}}$ is orthogonal to another, by counting the number of valid compositions $\vec{k}$, we identify the dimension of the PI  space as
\begin{equation}\label{eq:dpi}
    d_{\rm PI}:=\dim(\LL_{\PI}(\mathcal{H})) = {n+3 \choose 3}=: {\rm Te}(n+1),
\end{equation}
in terms of the Tetrahedral numbers ${\rm Te}(n)$.

\subsubsection{The Schur basis}\label{sec:pi_schur_basis}
The representation theory of $S_n$ (and its Schur-Weyl dual $\SU(d)$) is particularly well understood. A great wealth of details can be found in, e.g., Refs.~\cite{harrow2005applications,bacon2006efficient}, and we recall elements necessary to our analysis. In particular, we aim at understanding the basis induced by the irreducible decomposition of the action of $S_n$ and the dimensions entailed.

The basis for $\HH$ in which the action of $S_n$, in Eq.~\eqref{eq:action_Sn}, block-decomposes, is known as the \emph{Schur basis} $\{ \ket{\lambda, T_{\lm}, q_{\lm}} \}_{\lm,T_\lm,q_\lm}$. Here, the irreps of $S_n$ are labelled by partitions of the integer $n$ with at most $2$ parts: $\lm := (\lm_1, \lm_2)$ with $\lm_1 \leq \lm_2\in \mathbb{N}$ and $\lm_1+\lm_2=n$. 
Each partition can be associated to a Young diagram with $\lm_i$ boxes in its i-th row (here with $i\in [2]$).

The $T_\lm$, labelling a basis of the $S_n$-irrep, are in correspondence to Standard Young Tableaux of shape $\lm$: a filling of the Young diagram boxes with \emph{distinct} integers taking values in the set $[n]$ and arranged in strictly increasing order both row- and column-wise.
For a given $\lm$, we have 
\begin{equation}\label{eq:catalan}
    d_{\lm} := |\set{T_{\lm}}| = \frac{n!(\lm_1-\lm_2+1)!}{(\lm_1+1)! \lm_2!},
\end{equation} 
that is maximized (in the case of even $n=2l$) for $\lm=(l,l)$ with $\max_\lm d_{\lm} = n! / (l!(l+1)!)=:C(l)$ with $C(l)$ the $l$-th Catalan number. 

The $q_{\lm}$, labelling the multiplicities of the $S_n$-irrep, are in correspondence to Semi-Standard Young Tableaux: a filling of the Young diagram boxes with \emph{potentially repeated} integers taking values in the set $[2]$, and arranged in non-decreasing order row-wise and strictly increasing order column-wise. 
For a given $\lm$, we have 
\begin{equation}\label{eq:dim_q_reg}
    m_{\lm} := |\set{q_{\lm}}| = \lm_0 - \lm_1 +1,
\end{equation} 
that is maximized for $\lm=(n,0)$ with $\max_\lm m_{\lm} = n+1$. One can further verify that $\sum_{\lm} m_{\lm}^2 = {n+3 \choose 3}$ as would be expected from Eq.~\eqref{eq:dpi}.

\subsection{Deep permutation-invariant classical shadows}\label{sec:pi_protocol_1}

As detailed in Sec.~\ref{sec:cs_symm_rep}, provided known implementation of the block-diagonalization transform one can readily implements symmetric CSs. In our case, such transform is the Quantum Schur Transform (QST) for which several implementations have been proposed~\cite{bacon2005quantum,bacon2006efficient,kirby2017practical,krovi2019efficient}. These all require additional ancillas and a number of layers scaling at least linearly with the system size. 

According to Sec.~\ref{sec:cs_symm_rep}, we can readily instantiate two PI-CS protocols by performing the QST, applying a unitary onto the multiplicity register (here corresponding to the state $\ket{q_{\lm}}$)
drawn form either the LC or the GC ensembles
and measuring in the computational basis both the irrep and multiplicity registers. We denote as QST-LC and QST-GCs the corresponding PI-CS.

Inserting the dimension of the multiplicity register, from Eq.~\eqref{eq:dim_q_reg}, into  Eq.~\eqref{eq:var_RP_RC} we get  bounds on the variance:
\begin{align}\label{eq:var_RP_RC_PI}
\begin{split}    
    \var[\hat{o}_{\lm}]_{\rm QST-LC} & \leq (n^2 + 2n +2)||O||^2_\infty, \\
    \var[\hat{o}_{\lm}]_{\rm QST-GC} & \leq 3 (n^2 + 2n +2) ||O||^2_{\infty},
\end{split}
\end{align}
that shows quadratic scaling with $n$.

\subsection{Shallow permutation-invariant classical shadows}\label{sec:pi_protocol_2}

The previous already provides concrete PI-CS, but necessitates large circuit overhead.
Aiming for shallow symmetric CSs we consider the symmetrization approach advocated in Sec.~\ref{sec:symm_cs_symm}, and take as a starting point for symmetrization the non-symmetric CSs consisting of single-qubit rotations and computational basis measurements. The resulting PI-CSs, dubbed \emph{symm-PI} CSs, are described in the following and all necessary details for their implementations are provided in App.~\ref{app:mc_var_symm_pics} and~\ref{app:mc_schur_basis}.

\subsubsection{Symmetrized protocol}
A set of $3n$ generators for the products of single-qubit unitaries can be chosen as $\mathcal{G}=\set{X_i, Y_i, Z_i}$ for $i \in [n]$. Twirl of these $3n$ generators yields the set of $3$ PI generators $\mathcal{G}^{S_n}=\set{\sum_i X_i, \sum_i Y_i, \sum_i Z_i}$ generating
\begin{equation}\label{eq:V_PI}
    \VDIST_{\rm symm-PI} 
    =\set{ W^{\otimes n} \,|\, W \in \textsf{SU}(2)} 
\end{equation}
which is the same as the correlated local unitaries of Eq.~\eqref{eq:collective_V}.
Symmetrization of the computational basis measurements of Eq.~\eqref{eq:cb_meas} yields the measurement
\begin{equation}\label{eq:A_PI}
    \POVM_{\rm symm-PI} = \set{\pihw }_{h=0}^n
\end{equation}
of the $n+1$ \emph{Hamming-weight projectors} defined as
\begin{equation}\label{eq:hw_proj}
    \pihw:=\sum_{\substack{x \in \{0,1\}^{n}:\\ |x|=h}} \ket{x}\bra{x},
\end{equation}
where $|x|$ is the Hamming-weight (i.e., number of $1$s) of a bitstring $x$. 
In essence, the Hamming-weight projectors group all rank-one computational basis projectors $|x\rangle\langle x|$ that are related by permutations, breaking $\L$ into  equivalence classes labelled by Hamming weights $h$. 

Given an outcome $x\in \{0,1\}^n$ for a computational basis measurement, as obtained in typical quantum computing platforms, the outcome $h$ corresponding to our symmetric $\POVM_{\rm symm-PI}$ is computed by counting the bits of $x$ that are one. Note, however, that performing the measurement in Eq.~\eqref{eq:A_PI} does not require single particle resolution, something that could be desirable in the case of bulk quantum computing platforms~\cite{knill1998effective,knill2002introduction} and in some metrological setups~\cite{davis2016approaching,linnemann2016quantum}.

\subsubsection{Visible space}\label{sec:proof_vizspace}
Having defined symm-PI CSs,
we now verify that their visible space satisfies the symmetric CS criterion of Eq.~\eqref{eq:symm_CS_rule}. 
First, given a single-qubit operator $M\in \L(\mbb{C}^2)$, define for each $k\in[0,n]$ its symmetrized version
\begin{align}
\begin{split}
M^{(k)} := \TC_{S_n}(M^{\otimes k} I^{\otimes n-k}) \in \L.
\end{split}
\end{align}
In particular, we have $n+1$ $Z$-basis operators
\begin{align}
\begin{split}
Z^{(k)} = \TC_{S_n}(Z^{\otimes k} I^{\otimes n-k}).
\end{split}
\end{align}

One can show that the span of the set $\set{Z^{(k)}}_k$ is the same as the span of the set $\set{\pihw}_h$: they are both orthogonal bases of the $n+1$ dimensional subspace of PI operators diagonal in the computational basis. By linearity, it follows that the visible space of symm-PI can be expressed as
\begin{align}\label{eq:viz_ZU}
    \begin{split}
    \LVIS &:= \spann\{ (W^{\otimes n})^{\dag} \pihw W^{\otimes n}  \}_{h, W}\\
    &= \spann\{ (W^{\otimes n})^{\dag} Z^{(k)}  W^{\otimes n}\}_{k, W} \\
    &= \spann\{  Z_W^{(k)} \}_{k, W}.
    \end{split}
\end{align}
We defined $Z_W := W^{\dagger} Z W$. Note that there always exists a $W\in \SU(2)$ such that $Z_W = \alpha_X X + \alpha_Y Y + \alpha_Z Z$ for any $\vec{\alpha}=(\alpha_X, \alpha_Y, \alpha_Z)$ satisfying $||\vec{\alpha}||^2 = 1$. 

Finally, and central to the argument, we need to recall a result from Ref~\cite{harrow2013church} (Corollary. 4) stating that
\begin{equation}\label{eq:corro4}
\LPI = \spann \{ M^{\otimes n}\}_{M \in \LL(\mathbb{C}^{2})}.  
\end{equation}
That is, any PI operator can be written as a linear combination of $M^{\otimes n}$ where $M$ are arbitrary single-qubit observables. Without loss of generality, we can restrict them to be of the form $M = Z_W + \alpha_0 I$, for some $W\in \SU(2)$ and $\alpha_0 \in \mathbb{R}$.
Taking the $n$-fold tensor product of $M$, and grouping the resulting terms by numbers of $Z_W$ involved, we get 
\begin{align}\label{eq:full_vis_space}
\begin{split}
\LL^{\PI} &= \bigcup^n_{k=0} \; \spann \{ Z_W^{(k)}\}_W \\&= \spann \{ Z_W^{(k)}\}_{k,W} =  \LVIS 
\end{split}
\end{align}
with the last equality originating from Eq.~\eqref{eq:viz_ZU}.
The symm-PI CSs are thus symmetric CSs.

\begin{figure*}[t]
\centering
\includegraphics[width=1.95\columnwidth]{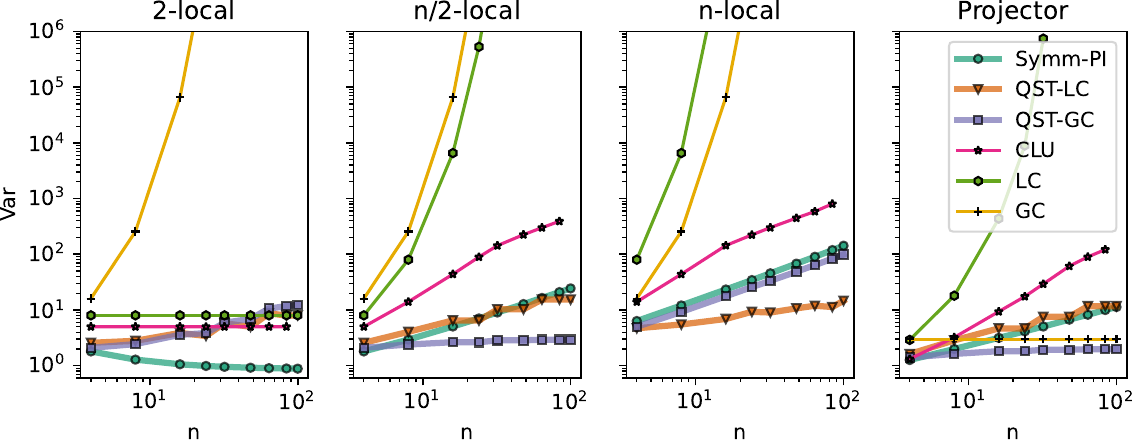}
\caption{
\textbf{The GHZ benchmark:} Single-shot variance for system sizes $n$ ranging from $4$ to $100$ qubits and representative observables including $Z_1Z_2$ (2-local), $Z^{\otimes n/2}$ ($n/2$-local), $Z^{\otimes n}$ ($n$-local) and  $|{\rm GHZ}\rangle \langle {\rm GHZ}|$ (projector). For each observable we compare the variances entailed by $3$ flavours of PI-CSs (highlighted colours and detailed in the main text) to more established but non-symmetric CS (discussed further in the main text).}
\label{fig:ghz_paulivar}
\end{figure*}

\subsubsection{Measurement channel and variance} 
Having verified that our symm-PI CSs satisfy Eq.~\eqref{eq:symm_CS_rule}, we proceed to compute their measurement channel. 
As per Eq.~\eqref{eq:mat_meas_channel} this channel is fully specified by a $d_{\PI} \times d_{\PI}$ matrix $C$ of weights coupling elements of $\LPI$ to another. 
In App.~\ref{app:mc_var_symm_pics} and~\ref{app:mc_schur_basis}, we derive expressions for those weights for both the symmetrized Pauli strings basis of Eq.~\eqref{eq:A_k} and the irrep operator basis of Eq.~\eqref{eq:basis_rep_th}. 
Despite closed-form expressions for the entries of the matrix $C$, characterizing the measurement channel in Eq.~\eqref{eq:mat_meas_channel}, we note that its inversion, as needed to evaluate Eq.~\eqref{eq:cs_obs}, has to be performed numerically. 

For the symmetrized Pauli strings basis, evaluation of the weights requires explicit integration over the representation $W^{\otimes 2n}$ of $W \in \SU(2)$ for arbitrary $n$, and we follow the approach of Ref.~\cite{van2022hardware}.
Numerical inspection reveals that the smallest eigenvalue of $C$ equals $1/(2n+1)$, which as detailed in App.~\ref{app:var_o_symmcs}, yields an upper bound
\begin{equation}\label{eq:bound_ev_M_PI}
    \var[\hat{o}]_{\rm symm-PI} \leq ||O||^2 (2n+1),
\end{equation}
on the variance. This already shows that for operators with constant norm (such as projectors), variance in the CS estimates can grow, at worst, linearly in the system size. This is in contrast to the exponential scaling entailed by LCs (but not GCs) as shown in Eq.~\eqref{eq:var_RP_RC}.
Still, the bound of Eq.~\eqref{eq:bound_ev_M_PI} can often be extremely loose for operators $O$ with norm $||O||$ scaling with $d$. 
We also provide expressions for evaluating exactly the variances  case by case, i.e., for a given operator $O$ and state $\rho$ (App.~\ref{app:ev_cs_variances}). 

In the irrep operator basis, evaluation of the entries of $\W$ involve Clebsch-Gordan coefficients that can be computed at scale~\cite{havlivcek2018quantum}. 
The merit of expressing $\MM$ in such basis is that the matrix $\W$ adopts a block diagonal structure with a total of $2n-1$ blocks.
Hence, inversion of $\W$ can be performed over smaller matrices, greatly easing this computational step\footnote{In practice, we never perform such matrix inversion. Rather, given that we always aim at computing quantities of the form $\W^{-1} \ket{o}$, we perform once a LU decomposition of $\W$, and use it later to solve equations of the form $\W \ket{x} = \ket{o}$. This has the same overall complexity as inverting $\W$, but is known to be more stable. In the case of block diagonal $\W$, the LU decomposition can also be performed independently on each of the blocks. }.

\subsection{The GHZ benchmark}\label{sec:pi_dem}

To demonstrate the benefits of the PI-CSs we focus on a GHZ benchmark often considered for assessing CSs performances and also in the context of tomography and certification~\cite{carrasquilla2019reconstructing,huang2020predicting,huang2024certifying}. 
The underlying state, assumed to be unknown, is a GHZ state $\ket{\Psi} = (\ket{0}^n + \ket{1}^n)/\sqrt{2}$ that allows us to probe sizeable systems (up to $n=100$ qubits) in numerical simulations such as to assess the scalability of the PI-CSs; we are tasked to estimate properties relevant to many quantum computing applications, namely expectation values of Pauli strings, with varied locality, and of projectors.

We compare PI-CSs to established but non-symmetric CSs. 
Non-symmetric CSs include the random Local Clifford (LC) and Glocal Clifford (GC) CSs~\cite{huang2020predicting}, with unitary ensembles defined in Eq.~\eqref{eq:RP_RC}, and also the correlated-local-unitary CSs (CLU)~\cite{van2022hardware} defined in Eq.~\eqref{eq:collective_V}. 
PI-CSs include the ones based on the QST followed by either random Local Clifford (QST-LC) or random Global Clifford (QST-GC) measurements, presented in Sec.~\ref{sec:pi_protocol_1}, and the ones obtained by symmetrization (symm-PI), presented in Sec.~\ref{sec:pi_protocol_2}.

\begin{table}
\centering
\begin{tabular}[t]{l cccccc}
\hline
Obs.& \textbf{Symm-PI} & \textbf{QST-LC} & \textbf{QST-GC}  & CLU & LC & GC \\
\hline
$Z_1Z_2$ & $\mathbf{ \frac{1}{log(n)} }$ & $\mathbf{\sqrt{n}}$ &  $\mathbf{\sqrt{n}}$ & $1$ & $1$ & $2^n$\\
$Z^{\otimes n/2}$& $\mathbf{n}$& $\mathbf{n}$& $\mathbf{1}$& $n$& $3^{\frac{n}{2}}$ & $2^n$\\
$Z^{\otimes n}$& $\mathbf{n}$ & $\mathbf{log(n)}$& $\mathbf{n}$& $n$& $3^n$ & $2^n$\\
$\ket{\Psi}\bra{\Psi}$ &$\mathbf{\sqrt{n}}$ &$\mathbf{\sqrt{n}}$ &$\mathbf{1}$& $n$ & $3^n$& $1$\\
\hline
\end{tabular}
\caption{Scalings of the variances for varied observables and CSs for the GHZ benchmark fitted to the data shown in Fig.~\ref{fig:ghz_paulivar}.}
\label{tab:ghz_scal}
\end{table}
In Fig.~\ref{fig:ghz_paulivar} we report single-shot variances, as defined in Eq.~\eqref{eq:var} and estimated empirically over $10^5$ samples. These variances tell us about the sampling required to achieve a given estimation error through Eq.~\eqref{eq:sampling_scaling} and are evaluated for a comprehensive list of observables:
From left to right we have included the observables $Z_1Z_2$, $Z^{\otimes (n/2)}$, $Z^{\otimes n}$ and the projector $\ket{\Psi}  \bra{\Psi}$. To facilitate the comparison, we provide in Tab.~\ref{tab:ghz_scal} the scalings of the variances with respect to the system size, extracted from the data of Fig.~\ref{fig:ghz_paulivar}.

For all the observables considered, we report a scaling in the variances of any of the PI-CSs \emph{at most linear} with the system size $n$. 
This is, in contrast with the tomographic-complete CSs that exhibit, at times, exponential scalings as expected through Eq.~\eqref{eq:var_RP_RC}. 
For instance, the RPs have constant variances for observables with fixed locality (such as $Z_1Z_2$) but exponentially increasing ones for all the other observables probed. 
Similarly, RCs are appropriate for projectors (such as $\ket{\Psi}\bra{\Psi}$) but yield exponential scalings for observables with norm $||O||=d$, as is the case for any Pauli string. 

Comparison between the different PI-CSs show that none of them  is systematically better than another.
For instance, Symm-CS is the best performing for the local observable $Z_1Z_2$, even showcasing decreasing variance, but performs equally well or worst on others.
Given the relatively modest advantage offered by QST-RPs and QST-RCS for only a subset of the observables, and the fact that they require substantially deeper circuits, we deem symm-CS to be the most appropriate.

Comparison between symm-PI to CLU shows that the former always perform better, with improvement ranging from constant factors improvements (c. $5$ to $10$ times smaller variances for $Z^{\otimes n}$ and $Z^{\otimes (n/2)}$), to better asymptotic scalings (e.g., a quadratic improvement for the projector $\ket{\Psi}  \bra{\Psi}$). 
Recall that the only difference between these two CSs lies in the measurements performed, and that from the computational basis measurements of CLU one can always infer results for the Hamming-weight measurements of symm-PI. 
This means that, from a CLU dataset one can always obtain a Symm-PI one. Hence, we have identified an instance where symmetry knowledge can be incorporated \emph{a posteriori}, by means of adequate classical post-processing, and still substantially improves estimates accuracy.

Finally, let us comment on the distinction between PI quantum tomography (PI-QT)~\cite{toth2010permutationally} and the PI-CS presented here. 
While PI-QT is also efficient, we highlight that it requires a number of measurements that has to scale quadratically with the system size~\cite{toth2010permutationally}. 
This is systematically larger than the scalings of the PI-CS reported in Fig.~\ref{fig:ghz_paulivar} (from decreasing to linear scalings) and guaranteed for the projectors in Eq.~\eqref{eq:bound_ev_M_PI} (at most linear). 
We also noticed in additional numerics that the accuracy of the PI-QT greatly depends on the specific choice of a fixed PI measurement basis. 
While one can optimize this choice through appropriate cost functions~\cite{toth2010permutationally}, we found such non-linear optimization to quickly become too demanding as the system size increases.

Overall, we confirmed the advantage of PI-CS compared to a wide range of established CSs and other tomography protocols.

\section{Permutation Invariant Classical Shadows in qudit systems}
In this final part we  explain how the PI-CS can be lifted to \emph{qudit} systems. After recalling some elements for the treatment of qudits (Sec.~\ref{sec:qudit_basis}) we detail how both the PI block-CS (Sec.~\ref{sec:pi_protocol_1_qudits}) and the PI symm-CS (Sec.~\ref{sec:pi_protocol_2_qudits}) are extended to this new setting.

\subsubsection{Qudit systems}\label{sec:qudit_basis}
Let $\HH^{(D)}:=(\mathbb{C}^{D})\tn = \spn \{ \ket{\vec{j}} \}_{\vec{j}\in [D]^n}$ be the Hilbert space of $n$ qudits (with dimension $D$).
In lieu of the Pauli operators, we can employ
the generalized 
Pauli matrices $Q_l$, which are products of shift and clock operators~\cite{gottesman1998fault}, as an orthogonal operator basis
for each qudit $\L(\mbb{C}^D) = \spn\{ Q_l\}_{l=1}^{D^2}$.
Similarly, the $n$-qudit operator space is spanned by the generalized Pauli strings $\L^{(D)}\coloneqq \LL(\HH^{(D)}) = \spn\{ Q_{\vec{l}} \}_{\vec{l}\in [D^2]^n}$ where $Q_{\vec{l}} = Q_{l_1} \otimes \hdots \otimes Q_{l_n}$.
To generalize the notion of Hamming-weight used in Eq.~\eqref{eq:hw_proj} and composition used in Eq.~\eqref{eq:A_k}, we introduce the concept of \emph{type}.
Given a string $x \in [s]^n$ composed of $n$ symbols, each chosen amongst $s$ values, we define its type $\vec{t}(x)$ as the $s$-tuple containing the number of times each of the $s$ symbols appears. We further denote as $T(n, s)$ the set of all valid types, which has cardinality $|T(n,s)|= {n+s-1 \choose s-1}$.
Given a string $x \in [s]^n$ composed of $n$ symbols, each chosen amongst $s$ values, we define its type $\vec{t}(x)$ as the $s$-tuple containing the number of times each of the $s$ symbols appears. We further denote as $T(n, s)$ the set of all valid types, which has cardinality $|T(n,s)|= {n+s-1 \choose s-1}$.

In essence, these types label equivalence classes under the action of $S_n$: Provided they have the same type, any string $x$ can be transformed to another string $x'$ through a permutation of its entries. As such, types naturally appear when performing twirls. For instance, twirling the generalized Pauli strings yields a basis for the PI operator subspace
\begin{equation}\label{eq:A_k_D}
\LL_{\rm PI}^{(D)}= \spn\{ A^{(D)}_{\vec{t}} = \TC_{S_n} (Q^{\vec{t}})\}_{\vec{t} \in T(n, D^2)} \subset \L^{(D)},
\end{equation}
where $Q^{\vec{t}} := Q^{\otimes t_1}_1 \otimes \hdots \otimes Q^{\otimes t_{D^2}}_{D^2}$
The basis of PI qudit operators of the form Eq.~\eqref{eq:A_k_D} generalizes the symmetrized Pauli strings for qubits defined in Eq.~\eqref{eq:A_k}. As before, one can verify that these operators are orthogonal, such that we can readily obtain the dimension of the PI qudit-operator space as $\dim(\LL_{\rm PI}^{(D)}) = {n+D^2-1 \choose D^2-1}  $. Similarly, twirling projectors onto the qudit computational basis 
we obtain symmetrized \emph{type} projectors 
\begin{equation}\label{eq:projectors_cb_qudits}
    \Pi_{\vec{t}}:=\sum_{\substack{x \in \{0,\hdots, D-1\}^{n}:\\ \vec{t}(x)=\vec{t}}} \ket{x}\bra{x}.
\end{equation}
generalizing the Hamming-weight projectors of Eq.~\eqref{eq:hw_proj}. 

Finally, we note that the representation theory of $S_n$ acting on qudits is similar to the one for qubits, with the notable exception that now the set of irreps is labelled by all partitions of $n$ with at most $D$ parts (rather than two).
Both the basis of the $S_n$ irreps and its multiplicity space are still in one-to-one corresponce with Standard  and Semi-Standard Young Tableaux respectively (as discussed in Sec.~\ref{sec:pi_schur_basis}). The latter now being a filling of Young diagram boxes with values in $[D]$ (instead of $[2]$), we get for the multiplicity of irrep $\lm$ in $\HC^{(D)}$,
\begin{equation}\label{eq:dim_q_reg_qudits}
    m_{\lm} = \frac{ \prod_{ 1\leq i < j \leq D } (\lm_i - \lm_j + j -i)}{\prod^D_{m=1} m!} \leq {n+D-1 \choose D-1},
\end{equation} 
such that $m:= \max_{\lm} m_{\lm} \in \mathcal{O}((n+D)^{D-1})$.

\subsubsection{Deep PI-CS through QST}\label{sec:pi_protocol_1_qudits}
To lift the PI-CS of Sec.~\ref{sec:pi_protocol_1} to qudits, it suffices to note that implementations of the QST extend to qudits~\cite{bacon2005quantum, bacon2006efficient,kirby2017practical,krovi2019efficient}. These implementations as quantum circuits are efficient in the sense that they require a number of gates that scales polynomially in both the system size $n$ and the qudit dimension $D$ (or $log(D)$ in Ref.~\cite{krovi2019efficient}).

Given the dimensions of the multiplicity space provided in Eq.~\eqref{eq:dim_q_reg_qudits}, the bounds on the variances for the qudit case are obtained, from Eq.~\eqref{eq:var_block_CS}, as
\begin{align}\label{eq:var_RP_RC_PI_qudits}
\begin{split}    
    \var[\hat{o}_{\lm}]_{\rm QST(D)} & \leq \mathcal{O}\left((n + D)^{2D} ||O||^2_\infty \right), 
\end{split}
\end{align}
This is to be compared to bounds of the order $D^{2n} ||O||^2_\infty$ for (non-symmetric) tomographic-complete CSs.

\subsubsection{Shallow PI-CS through symmetrization}\label{sec:pi_protocol_2_qudits}
A symmetrized qudit PI-CS protocol can be derived following the steps of Sec.~\ref{sec:pi_protocol_2}. Starting with product of random unitaries, the symmetrized ensemble becomes
\begin{equation}\label{eq:V_PI_qudits}
    \VDIST^{(D)}_{\rm symm-PI} =\set{ V=W^{\otimes n} \, | \, W \in \SU(D)},
\end{equation} 
consisting in tensor products of the same random unitary on each qudit. Parametrization of $\SU(D)$ unitaries can be achieved through generalized Euler angles~\cite{tilma2002generalized}.
Symmetrizing computational basis measurements yields
\begin{equation}\label{eq:A_PI-qudits}
    \POVM^{(D)}_{\rm symm-PI} = \set{ \Pi_{\vec{t}} }_{\vec{t} \in T(n, D)}
\end{equation}
consisting in the type projectors  defined in Eq.\eqref{eq:projectors_cb_qudits}.

The unitary ensemble and measurements, in Eqs.~\eqref{eq:V_PI_qudits} and~\eqref{eq:projectors_cb_qudits} respectively, define the symmetrized PI-CS for qudits. 
Proof that this is indeed a symmetric CS, satisfying Eq.~\eqref{eq:symm_CS_rule}, follows from the fact that Eq.~\eqref{eq:corro4}  is equally valid in the qudit setting, such that
\begin{equation}\label{eq:corro4_qudits}
\LL_{{\rm PI}}^{(D)} = \spann \{ M^{\otimes n}\}_{M \in \LL(\mathbb{C}^{D})}.  
\end{equation}
In App.~\ref{app:mc_var_symm_pics} and~\ref{app:mc_schur_basis}, we also sketch how expressions of the measurement channel for this symmetrized qudit PI-CS could be derived following the methodology used to obtain the qubit one. Altogether, these constitute the necessary elements to perform PI-CS on qudits with constant (independent of $n$, but scaling with $D$) circuit overhead.

\section{Conclusion and Outlook}

\medskip
\noindent
In this work, we departed from the most often considered situation in the CS literature where one assumes no knowledge regarding the underlying state or the observables involved, and which thus necessitates tomographic-complete protocols. Instead, we focused on scenarios where a priori knowledge is available, as \textit{symmetries} of the unknown state or of the properties to be probed. 
In such cases, we argued for the systematic imposition of such symmetries in the construction of the CS protocols, and proposed simple criterion for \textit{symmetric CS}.
Focusing on the symmetric group, and building upon previous works, we studied several realizations of PI-CSs, that were shown to consistently improve on other established but more generic CSs.
While mostly discussed in the context of qubit systems we also shown how the protocols presented could readily be extended to qudits.

\subsection{Applications of PI-CS}
CSs are ideally positioned as a central tool in the benchmarking of quantum systems~\cite{huang2024certifying}.
The PI-CS developed here are naturally adequate for the certification and calibration of resource states including GHZ states (as studied in the numerical demonstration section) and families of spin squeezed states~\cite{wineland1992spin,kitagawa1993squeezed} core to many metrology protocols~\cite{davis2016approaching,linnemann2016quantum,kaubruegger2019variational,volkoff2022asymptotic}.
Going further, PI-CS can support 
more general benchmarking suites.
Its sampling efficiency combined with the classical simulability of preparation of PI states through PI circuits~\cite{anschuetz2022efficient,goh2023lie} enable scalability.
Furthermore, except in dedicated platforms, implementations of PI quantum operations would require compilation of such operations in terms of many non-PI primitive gates.
Hence, PI-CS could form the basis of scalable, and still very demanding, benchmarks on which to assess quantum platforms.

In addition to metrology, forms of PI naturally occur in several quantum information tasks, for which PI-CS could present an interesting addition. 
This is the case, in problems involving learning over multiple copies of a state\footnote{Given that the ordering of the copies is irrelevant, an optimal protocol for such tasks can always be made PI with respect to permutation of the copies.}, such as spectrum estimation and tomography.
In these problems, optimal protocols rely on measurements performed through the use of the QST~\cite{alicki1988symmetry,keyl2001estimating,keyl2006quantum,haah2016sample,o2016efficient,o2017efficient}.
Given the connections between PI and the QST, one could wonder whether the lightweight symmetrized PI-CS detailed here could be used while still retaining some of the advantages of the QST-based ones. 
Also, in more synthetic problems of quantum machine learning on graphs~\cite{mills2019quantum,verdon2019quantumgraph,skolik2022equivariant,schatzki2022theoretical,mernyei2022equivariant,albrecht2023quantum,umeano2024geometric,dalyac2024graph}, whereby PI properties are measured on quantum states encoding the underlying graphs, 
we expect that PI-CS could expand capabilities of current models.

\subsection{Future works}
First, let us comment on the power of symmetries to prescribe families of states or observables.
In this work, as per Eq.~\eqref{eq:linear_symm}, symmetries were defined as groups of operators commuting with the states or observables of interest. 
We should recognize that there are situations where such definition of symmetries is too narrow and may not capture the structure of the underlying states. 
As an illustration, consider the case when the state $\rho$ is known to be a product state. 
Despite great reduction in the degrees of freedom, from exponential to linear, the only symmetries present would be the trivial one.
A similar situation arises if $\rho$ is known to be a free-fermionic state, an information which cannot be described by a priori known symmetries.
In both cases, however, the structure of the underlying families of states is better captured through the use of quadratic symmetries that are the group of operators that commute with $\rho^{\otimes 2}$~\cite{zimboras2015symmetry}.
It will be of great interest to explore how the concepts presented here extend to quadratic symmetries.

Second, we note that since its inception the field of CSs has quickly expanded; we expect that many of the recent developments could be incorporated to the symmetric CSs, such as the inclusion of built-in error-mitigation strategies~\cite{chen2021robust,koh2022classical,zhao2023group}, the exploration of bias-variance trade-offs~\cite{van2022hardware,cai2024biased}, and extensions to quantum channels~\cite{caro2022learning,huang2023learning,jerbi2023shadows}. 

More generally, we hope that this work will stimulate the development and studies of many more symmetric CSs (i.e., for other groups of symmetries or actions). 
As a natural group of symmetries to consider, $\SU(D)$ occurs in many physical situations. 
Already, due to the Schur-Weyl duality $\SU(D)$ symmetric CSs follows from Sec.~\ref{sec:pi_protocol_1} when swapping the roles of the multiplicity and dimension registers. In this case, through Eq.~\eqref{eq:catalan}, we can compute a reduction of the upper bound in Eq.~\eqref{eq:var_RP_RC_PI} of the order of $n^3$ compared to tomographic-complete CSs.
However, as relying on the QST it is deemed too demanding for near-term implementations. 
This motivates the search for shallower $\SU(D)$ symmetric CSs as may be possible through Sec.~\ref{sec:symm_cs_symm}.

As quantum technologies mature, however, the use of deeper circuits will become more adequate. This will encourage the development of new block diagonalization transforms, akin to the QST. 
Many more symmetries such as subgroup of the symmetric groups, spatial symmetries that have been found to be beneficial when designing parameterized quantum circuits~\cite{sauvage2022building}, or more exotic ones as found in chemistry problems and high-energy physics models would represent interesting research directions.

Ultimately, adoption of specific symmetric CSs will depend on trade-offs between reduction in sampling complexity and circuit overheads entailed. These will depend on the nature and dimension of the group of symmetries of interest and also on explicit details of the protocols. As discussed earlier, symmetric CS are not unique, such that exploration of many symmetric CSs is envisioned.

\medskip
\section{Acknowledgements}
We thank Etienne Granet and Nathan Fitzpatrick for useful comments on the manuscript. M.L.
was supported by the Center for Nonlinear Studies at LANL and by the Laboratory Directed Research and Development (LDRD) program of LANL under project number 20230049DR.

\clearpage
\onecolumngrid

\clearpage
\appendix
For convenience, we start these appendices by recalling (and extending on) some notations and background from the main text (Sec.~\ref{app:background}) and we providing information regarding the symmetric group and permutation-invariant (PI) operators (Sec.~\ref{app:bck_sn}). 
We then present necessary details for the implementation of the symmetrized PI-CS in the symmetrized Pauli basis (Sec.~\ref{app:mc_var_symm_pics}). These include, expressions for computation of the measurement channel and resulting variances. This is completed with additional details for the implementation of the symmetrized PI-CS in the Schur basis (Sec.~\ref{app:mc_schur_basis}). The latter has the merit of reducing the computational burden entailed by the inversion of the measurement channel.

\section{Background}\label{app:background}

After recalling notation  (Sec.~\ref{app:bck_notations}), we review background material regarding CSs (Sec.~\ref{app:bck_cs}), symmetries (Sec.~\ref{app:bck_symm}) and symmetrized CSs (Sec.~\ref{app:bck_symm_cs}). 

\subsection{Notation}\label{app:bck_notations}
Given a vector space $V$, let $\LL(V)$ be the space of linear operators acting on $V$ that includes $\BB(V)$ the subspace of Hermitian operators (also called observables) $A$ satisfying $A^{\dagger}=A$.
Unitaries on $V$ are the operators $U$ that satisfy $U^{\dagger}U=UU^{\dagger}=I_{V}$, with $I_{V}$ the identity on $V$.
They form a group identified through the dimension $d=dim(V)$ of $V$ as
$\textsf{U}(d)$. The special unitaries are unitaries with unit determinant, which group is denoted as $\textsf{SU}(d)$. 
Operator spaces are endowed with the Hilbert-Schmidt inner product $\langle A, B\rangle := \trace[A^{\dag} B]$ that induces the (Froebenius) norm $||A||:=\sqrt{ \trace[A^{\dag} A]}$. Sometimes, we will also make use of the spectral norm $||A||_{\infty}$ which for an observable $A \in \BB(\HH)$ is simply its largest absolute eigenvalue.

The systems of interest are $n$-qudit Hilbert spaces $\mathcal{H}:=(\mathbb{C}^{D})^{\otimes n}$ with $D$ the dimension of a single qudit and $d=D^n$ the dimension of the overall system. 
Unless otherwise stated, we will focus on qubits system whereby $D=2$, $\HH = (\mathbb{C}^{2})^{\otimes n}$ and $d=2^n$ and for ease of notations we drop the $\HH$ in our notations. E.g., $\mathcal{L}$ and $\mathcal{B}$ denote the operators and observables acting on the $n$-qubit Hilbert space.
The set of single-qubit Pauli operators (including the identity) is denoted $\set{X, Y, Z, I}$. Pauli strings $P\in \{X, Y, Z, I\}^{\otimes n}$ are $n$-fold tensor of single-qubit Pauli operators. 
We call the composition of a Pauli string the vector $\vec{k}=(k_X, k_Y, k_Z, k_I) \in \mathbb{Z}^4$, where each $k_\alpha$ is the number of times a given Pauli operator $\alpha \in \set{X, Y, Z, I}$) appears in the string. For instance the Pauli string $P=XYXZI$ has composition $\vec{k}=(2,1,1,1)$. By construction, the composition $\vec{k}$ of a $n$-qubit Pauli string has to satisfy $|\vec{k}|:=k_X+k_Y+k_Z+k_I=n$.

In tasks of CSs, for an unknown state n-qubit state $\rho$ one is tasked to estimate expectation values $o_m := \trace[\rho O_m]$ for a set of $M$ observables $\set{O_m}$ with $m \in [M]:=\set{1, \hdots, M}$. 
Estimates are indicated by hats with, e.g., $\hat{o}_m$ being an estimate of $o_m$. Finally, we denote expectation values of a function $f$ given a probability distribution $\prob(X)$ as $\E_{X}[f]$, and correspondingly $\E_{X,Y}[f]$ for expectation values under the joint distribution $\prob(X,Y)$.

\subsection{Classical Shadows}\label{app:bck_cs}
A CS protocol is specified in terms of a random unitary ensemble $\VDIST=(H,\mu)$ together with a fixed measurement $\POVM=\set{\Pi_w}$. The unitary ensemble consists in
a choice of $H\subset \textsf{SU}(d)$, a subgroup of the special unitaries, and $\mu$ a probability measure $\mu:H \xrightarrow[]{} [0,1]$. In this work, $\mu$ is always taken to be the uniform measure or Haar measure.
The measurement $\mathcal{A}=\set{\Pi_w}$ consists in a set of $\Pi_w$ that are projectors, with $\Pi_w \Pi_w = \Pi_w = \Pi_w^\dag$, and which satisfy $\sum_w \Pi_w = I$. Each random measurement performed consists on drawing an unitary $V$ with probability $\mu(V)$ and performing a measurement $\POVM$ that yields outcome $w$ with probability $\trace [V^\dag \Pi_w V \rho]$. The probability of jointly choosing $V$ and obtaining an outcome $w$ is $p_\rho(V,w)=\mu(V)\trace [V^\dag \Pi_w V \rho]$ and expectation values taken under such distribution are denoted $\E_{V,w}$. Expectation values taken under the random choice of unitaries only are denoted $\E_V$.

The measurement channel $\MM: \BB \mapsto \BB$ associated to the ensemble $\VDIST$ and measurement $\POVM$ is defined as
\begin{align}\label{eq:app_mc}
\begin{split}
\MM( \rho ) &:= \E_{V,w} \bigg[V^{\dagger} \Pi_w V \bigg]
     = \sum_w \int \mu(V) \trace[V^{\dagger} \Pi_w V\rho] (V^{\dagger} \Pi_w V)
     \\&=\trace_A \bigg[ \Big(\E_{V} \big[ (V^{\dagger})^{\otimes 2} \, \widetilde{\Pi} \, V^{\otimes 2}\big] \Big) \Big(\rho \otimes I\Big) \bigg], \;\; \text{with} \quad \widetilde{\Pi}:=\sum_{w} (\Pi_w)^{\otimes 2}.
\end{split}
\end{align}
In the last line $\trace_A[\cdot]$ indicates that we trace out the last $n$ qubits of the overall $2n$-qubit system.  
One can verify that the measurement channel is self-adjoint:
\begin{align*}\label{eq:app_mc_sa}
\begin{split}
     \langle X, \MM(Y) \rangle &= \trace \left[X^\dag \MM( Y ) \right] = \sum_w \int_{V} \mu(V) \trace[V^{\dagger} \Pi_w V Y] \trace [ X^\dagger V^{\dagger} \Pi_w V] \\ 
     &= \trace \left[ \left( \sum_w \int_{V} \mu(V) V^{\dagger} \Pi_w V  \trace [ X V^{\dagger} \Pi_w V]\right)^\dag Y\right] =  \trace \left[\MM(X)^\dag  Y \right] =: \langle \MM(X), Y \rangle.
\end{split}
\end{align*}
In the second line we used the identity $\trace[A]\trace[B] = \trace[A \otimes B]$.
This implies that $\mathcal{M}^{-1}$ is also self-adjoint: $\trace[X^\dag \MM^{-1}(Y)]=\trace[\MM^{-1}(X)^\dagger Y]$. When $\mathcal{M}$ is not invertible, we take $\mathcal{M}^{-1}$ to be its pseudo-inverse.

For a choice of unitary $V$ and measurement outcome $w$, a CS estimate of $\rho$ is obtained as 
\begin{equation}
    \hat{\rho}:= \mathcal{M}^{-1}(V^{\dagger} \Pi_w V)
\end{equation}
As shown in the main text, this estimate is unbiased as $\E_{V, w}[\hat{\rho}]=\rho$.
Given $o:=\trace[O \rho]$, its CS estimate defined as 
\begin{equation}\label{eq:app_cs_est_exp}
    \hat{o}:= \trace \left[ O \hat{\rho}\right] =  \trace \left[ O \MM^{-1}(V^{\dagger} \Pi_w V) \right] =  \trace \left[ \MM^{-1}(O) V^{\dagger} \Pi_w V \right]
\end{equation}
is, by linearity, also unbiased. Its variance, that depends on the state $\rho$ is bounded through
\begin{align}\label{eq:app_var}
\begin{split}
     \var[\hat{o}] &:= \E_{V,w}[\hat{o}^2] - o^2
     = \left( \sum_w \int_{V} \mu(V) \trace[V^{\dagger} \Pi_w V\rho]\trace[V^{\dagger} \Pi_w V \MM^{-1}(O)]^2 \right)- o^2 \\
     &\leq \left( \sum_w \int_{V} \mu(V) \trace[V^{\dagger} \Pi_w V\rho]\trace[V^{\dagger} \Pi_w V \MM^{-1}(O)]^2 \right)\\     
     &= \trace \bigg[ \Big(\E_{V} \big[ (V^{\dagger})^{\otimes 3} \, \doublewidetilde{\Pi} \, V^{\otimes 3}\big] \Big) \Big(\rho \otimes \MM^{-1}(O) \otimes \MM^{-1}(O) \Big) \bigg],\;\; \text{with} \quad \doublewidetilde{\Pi}:=\sum_{w} \Pi_w^{\otimes 3} .
\end{split}
\end{align}

Given that $\MM$ is self-conjugate it admits a set of eigenvectors $\set{E_i}$ that form an orthonormal basis of $\mathcal{B}$. The corresponding eigenvalues $\lm_i$ are real and taken in ascending order: $\lambda_1, \leq \hdots \leq \lambda_{d^2}$. As $\MM(E_i)=\lambda_i E_i$  we have:
\begin{equation}\label{eq:app_lambda_i_alt}
    \lambda_i = \trace[\MM(E_i) E_i] = \trace \left[\left( \sum_w \int_{V} \mu(V) \trace[V^{\dagger} \Pi_w V E_i] V^{\dagger} \Pi_w V \right) E_i \right] = \sum_w \int_{V} \mu(V) \trace[V^{\dagger} \Pi_w V E_i]^2,
\end{equation}
where we made use of the definition of $\mathcal{M}$ in Eq.~\eqref{eq:app_mc}. This shows that $\lm_i\geq 0$.  Furthermore, recall that the pseudo-inverse $\MM^{-1}$ is only defined on the image of $\MM$. 
That is, $\MM^{-1}(E_i) = \lambda_i^{-1} E_i$ for $\lm_i \neq 0$ and $\MM^{-1}(E_j) = 0$ otherwise.

We proceed by bounding the variance of the CS estimates $\hat{e}_i$ of $e_i:= \trace[E_i \rho]$. From Eq.~\eqref{eq:app_var} we get
\begin{align}\label{eq:app_var_ei}
\begin{split}
     \var[\hat{e}_i] &\leq \left( \sum_w \int_{V} \mu(V) \trace[V^{\dagger} \Pi_w V\rho]\trace[V^{\dagger} \Pi_w V \MM^{-1}(E_i)]^2 \right) \\
     &\leq \lambda_i^{-2} \left( \sum_w \int_{V} \mu(V) \trace[V^{\dagger} \Pi_w V E_i]^2 \right) = \lambda_i^{-1}.
\end{split}
\end{align}
where we have used that $\trace[V^{\dagger} \Pi_w V\rho] \leq 1$ (as it corresponds to a measurement probability) and $\MM^{-1}(E_i) = \lm_i^{-1}E_i$ to obtain the second inequality. For the last inequality, we inserted Eq.~\eqref{eq:app_lambda_i_alt}. Finally, given that $\set{E_i}$ forms an orthonormal basis, any observable $O$ can be decomposed as $\sum_i w_i E_i$ with $\sum_i w^2_i = ||O||^2$, such that
\begin{align}\label{eq:app_var_o}
\begin{split}
     \var[\hat{o}] &= \sum_i w_i^2 \var[\hat{e}_i] \leq \sum_i w^2_i (\lambda_i^{-1}) \leq \lm_0^{-1} ||O||^2.
\end{split}
\end{align}
where we used Eq.~\eqref{eq:app_var_ei} and $\lm_0^{-1} \geq \lm_i^{-1}$. The upper bound in Eq.\eqref{eq:app_var_o}, albeit in general not tight, has the merit of relating variance and eigenspectrum of the measurement channel in general situations.

\subsection{Symmetries}\label{app:bck_symm}
Let $G$ be group of symmetry and $U:G \mapsto \textsf{U}(d)$ a unitary representation on the Hilbert space $\mathcal{H}$ of dimension $d$. By definition, it satisfies 
$U(g)U(g')=U(g g')$ for any $g, g' \in G$. 
Any $g \in G$ acts on pure states $\ket{\psi} \in \HH$ as $g \cdot \ket{\psi} = U(g) \ket{\psi}$, and on linear operators $A \in \mathcal{L}(\mathcal{H})$ through conjugation $g \cdot A = U(g) A U^{\dagger}(g)$. We use the shorthand notation $U_g := U(g)$. 

An operator $A \in \mathcal{L}$ is said to be \emph{symmetric} whenever $[A, U_g]=0$, or equivalently whenever $U_g A U^{\dagger}_g=A$  (i.e., $A$ is invariant under the action of $g$), for all $g \in G$. 
The subspace of symmetric operators is defined as
\begin{equation}\label{app:linear_symm}
    \LG:=\set{A \in \mathcal{L} \,| \, [A, U_g]=0, \; \forall g \in G}.
\end{equation}
and is denoted as $\BB^{G}:=\BB \cap \LG$ when restricted to the Hermitian operators.

Given a group $G$, we can define the twirl as an average of the action of each group elements. When acting on an operator $A$, it is thus given by
\begin{equation}\label{eq:twirl}
    \mathcal{T}_G(A) := \frac{1}{|G|} \sum_{g \in G} U_gAU_g^{\dagger},
\end{equation}
and we often denote $A^{G}:=\mathcal{T}_G(A)$.
The twirl of Eq.~\eqref{eq:twirl} defined for a discrete group $G$ with cardinality $|G|$, can be extended to continuous compact group when replacing the sum by an integral with respect to the uniform distribution over $G$ (i.e., under the Haar measure). 

Given the group nature of $G$, for any $h\in G$ we have $hG :=\set{hg | g \in G} = G$, such that for any operator $A$,
\begin{equation}\label{eq:average_group}
    \sum_{g} U_h U_g A U^{\dagger}_g U^{\dagger}_h = \sum_{g} U_g A U^{\dagger}_g.
\end{equation}
Using Eq.~\eqref{eq:average_group} one can verify that (i) $A^G \in \LG$ for any operator $A$, (ii) $A^G = A$ for any $A \in \LG$ and that (iii) $(A^G)^G = A^G$. That is $\mathcal{T}$ is an orthogonal projector onto $\LG$ (or $\mathcal{B}^{G}$ when restricted to observables). Also from Eq.~\eqref{eq:average_group} and the cyclicity of the trace we can verify that
\begin{equation}\label{eq:app_equiv_expectations}
    \trace[A^G B] = \trace[A B^G] = \trace[A^G B^G],
\end{equation}
as advertised in Eq.~\eqref{eq:equiv_expectations} of the main text.

\subsection{Measurement channel and variance for symmetrized CSs}\label{app:bck_symm_cs}
Recall that, as per Sec.~\ref{sec:symm_cs_symm}, symmetrized CSs have, by construction, both symmetric random unitaries ensemble $\VDIST^G$ and symmetric measurements $\POVM^G$. That is for any unitary $V \in \VV^G$, projector $\Pi_w \in \POVM^G$ and symmetry $g\in G$ we have $[V, U_g]=[\Pi_w, U_g]=0$. We denote as $\MM_{\rm symm}$ the measurement channel of such symmetrized CS.
In the following we provide general expressions for the measurement channels and variances in terms of an orthonormal basis $\set{B_k}_{k\in [d_{G}]}$ of the symmetric observable space $\HERMG$ that has dimension $d_{G}:=\dim (\HERMG)$. Later on we will evaluate these expressions in the PI case where $G=S_n$.

For the measurement channel, we first note that the terms $(V^{\dagger})^{\otimes 2} \, \widetilde{\Pi} \, V^{\otimes 2}$ defined in Eq.~\eqref{eq:app_mc} are invariant under the group $G \times G$ with representation $U_{(g, h)} = U_{g}\otimes U_{h}$, for any $(g,h)\in G \times G$.
It follows that any combinations of such operators is invariant under $G \times G$ and decomposes in terms of $B_k \otimes B_{k'}$. In particular, 
\begin{align}\label{eq:twirl_pitilde}
    \E_{V} \bigg[ (V^{\dagger})^{\otimes 2} \, \widetilde{\Pi} \, V^{\otimes 2} \bigg] = \sum_{k, k' =1}^{d_G} c(k,k') B_{k} \otimes B_{k'}.
\end{align}
From Eq.~\eqref{eq:app_mc}, we then see that the measurement channel of the symmetrized CS adopts the form
\begin{align}\label{eq:mc_symm}
\begin{split}
     \MM_{\rm symm}(\rho) &= \sum_{k, k'} c(k,k') \trace[\rho B_{k}] B_{k'},
\end{split}
\end{align}
which is entirely determined by the coefficients $c(k,k')$ arising from the decomposition in Eq.~\eqref{eq:twirl_pitilde}.
As expected, only the symmetric components $\rho_k:=\trace[\rho B_{k}]$ of the input contribute, and the channel's image lies in the symmetric operator subspace. 
Given an operator $A \in \LL$ and $A^G=\TT(A)$ its symmetric restriction, let us define the $d_{G}$-dimensional vector $\ket{A^G}\rangle$ which has entries $\ket{A^G} \rangle_k=\trace[B^{\dag}_k A]$ for $k\in[d_G]$.
We can further pack the coefficients $c(k,k')$ into a $d_G \times d_G$ matrix $C$. Then, the action of the channel on an operator $A$ specified in Eq.~\eqref{eq:mc_symm} can be rewritten in vectorized form as:
\begin{equation}
\ket{\MM_{\rm symm}(A)}\rangle = \W \ket{A^G} \rangle.
\end{equation}

The pseudo-inverse $\mathcal{M}_{\rm symm}^{-1}$ is defined as the inverse of $\mathcal{M}_{\rm symm}$ on $\HERMG$ and $0$ otherwise. In vectorized form, its action on an operator $A$ is obtained through inversion of $\W$, as
$\ket{\MM_{\rm symm}^{-1}(A)}\rangle = \W^{-1} \ket{A^G}\rangle$. We note that non-invertibility of $\W$ is symptomatic of cases when the visible space is a strict subspace of the symmetric operator space. In cases of symmetric CSs, as per the definition Eq.~\eqref{eq:symm_CS_rule}, $\W$ is guaranteed to be invertible.

Evaluation of the CS estimates of expectation values, as defined in Eq.~\eqref{eq:app_cs_est_exp}, are given by 
\begin{equation}\label{eq:vec_cs_est}
    \hat{o}:= \trace \left[ O \hat{\rho}\right] = \langle \bra{V^{\dagger}\Pi_w V} \W^{-1} \ket{O^G}\rangle.
\end{equation}
To evaluate these expressions, and as noted in the main text, we never need exact details of $\W^{-1}$ but rather to be able to compute quantities such as $\ \W^{-1} \ket{O^G}\rangle$, or alternatively $\langle \bra{V^{\dagger}\Pi_w V} \W^{-1}$ when the number $M$ of observables is larger than the number $S$ of measurements. Hence, rather than performing inversion of $\W$ we can perform \emph{once} a LU decomposition of $\W$, and use it to solve the system of equations $\W \ket{x}\rangle = \ket{O^\TT}\rangle$. Such approach has the same complexity as inverting $\W$ and is known to be more stable.

For the variance defined in Eq.~\eqref{eq:app_var}, one can show that
\begin{equation}\label{eq:twirl_pi_triple}
\E_{V} \big[ (V^{\dagger})^{\otimes 3} \, \doublewidetilde{\Pi} \, V^{\otimes 3}\big] = \sum_{k,k',k''=1}^{d_G} c(k,k',k'') B_k \otimes    B_k' \otimes    B_k''  
\end{equation}
as any of the terms $(V^{\dagger})^{\otimes 3} \, \widetilde{\Pi} \, V^{\otimes 3}$ is invariant under $G \times G \times G \times G$ with representation $U_{(f, g, h)} = U_{f}\otimes U_{g}\otimes U_{h}$, for any $(f,g,h)\in G \times G$. When inserted into Eq.~\eqref{eq:app_var}, we get
\begin{align}\label{eq:var_symm}
\begin{split}
     \var_{\rm symm}[\hat{o}] &= \sum_{k', k''} c(k, k', k'') \rho_k \tilde{o}_{k'} \tilde{o}_{k''}
\end{split}
\end{align}
where $\rho_k := \trace [\rho B_k ]$ and $\tilde{o}_k := \trace [\mathcal{M}_{\rm symm}^{-1}(O) B_k ]$. 

In essence the coefficients $c(k,k')$ appearing in Eq.~\eqref{eq:twirl_pitilde} are key to the practical  use of symmetrized CSs, while the coefficients $c(k,k',k'')$ appearing in Eq.~\eqref{eq:twirl_pi_triple} are key in evaluating the variances entailed.
Later on, when studying concrete symmetrized CSs for the group of symmetry $G=S_n$ we will provide expressions to  evaluate them.

\section{Symmetric group, permutation invariance and bases for permutation-invariant operators.}\label{app:bck_sn}
We recall definition of the symmetric group $S_n$ and its action on states and operators (Sec.~\ref{app:symmetric_group}).  Then we detail two basis for the permutation-invariant (PI) operators including the symmetrized Pauli strings (Sec.~\ref{app:symmetric_group_spb}) and the PI Schur operator basis (Sec.~\ref{app:symmetric_group_opschur}). Subsequently, elements for conversion between these basis are presented (Sec.~\ref{app:symmetric_group_change}), and
we conclude by recalling the PI-CS protocol obtained through symmetrization (Sec.~\ref{app:bck_symm_pics}). 

\subsection{The Symmetric group}\label{app:symmetric_group}
The symmetric group $S_n$ is naturally thought of as the set of all the permutations over a set $[n]:=\set{1, \hdots, n}$ of $n$ indices. Given a permutation $\sigma \in S_n$, we denote its action on an index $i \in [n]$ as $\sigma(i)$, and recall that for any $\sigma \in S_n$, its inverse $\sigma^{-1}$ is also a permutation. Action of a permutation $\sigma \in S_n$ on any n-qubit separable state is defined through
\begin{equation}\label{eq:app_act_sn}
    U_{\sigma} \big( \ket{i_1} \otimes \hdots \otimes \ket{i_n} \big) = \ket{i_{\sigma^{-1}(1)}} \otimes \hdots \otimes \ket{i_{\sigma^{-1}(n)}}.
\end{equation}
That is, a permutation acts as a relabelling of the qubit indices.
This is extended by linearity to arbitrary states $\ket{\psi} \in \mathcal{H}$. 
In the computational basis, the unitaries $U_{\sigma}$ are real and sparse (a single $1$ per row and column and $0$ otherwise).
The action of $\sigma \in G$ on an operator $A$ is given by:
\begin{equation}\label{eq:app_act_sn_operator}
\sigma \cdot A:=U_{\sigma}AU_{\sigma}^{\dagger}.
\end{equation}
We say that an operator is permutation invariant (PI) whenever it is invariant under $S_n$, and denote the space of all PI observables as
\begin{equation}
    \mathcal{B}^{\PI} := \set{A \in \mathcal{B} \, | \, [U_{\sigma}, A]=0, \, \forall \sigma \in S_n}.
\end{equation}
Two bases for this space are now detailed.

\subsection{The symmetrized Pauli basis}\label{app:symmetric_group_spb}
As discussed in the main text, a basis of the symmetric space can be obtained by twirling a basis of the operator space.
For the latter we can choose the $4^n$ Pauli strings. As per Eq.~\eqref{eq:app_act_sn_operator}, the action of the symmetric group on a Pauli string only shuffles the order of the individual Pauli operators but does not affect their \emph{composition}. Recall that compositions of Pauli strings were defined in Sec.~\ref{app:bck_notations} and denoted through a vector $\vec{k}=(k_X, k_Y, k_Z, k_I)\in \mathbb{N}^4$ satisfying $|\vec{k}|=n$. Hence, under the twirl of Eq.~\eqref{eq:twirl}, any two Pauli strings having the same composition map to the same symmetrized operator.
Furthermore each set of Pauli strings with the same composition (i.e., each equivalence class under the action of $S_n$) can be represented by an ordered Pauli string.
It follows that an \emph{orthonormal} basis of $\piopspace$ is formed by the Hermitian operators (equivalent to Eq.~\eqref{eq:A_k} of the main text up to normalization)
\begin{equation}\label{eq:app_basis_piop}
    B_{\rm Pauli}=\Bigr\{ \piop := \frac{1}{ \sqrt{d n! k_X! k_Y! k_Z! k_I! }}\sum_{\sigma \in Sn}  U_\sigma \left( X^{\otimes k_X} Y^{\otimes k_Y} Z^{\otimes k_Z} I^{\otimes k_I} \right) U_\sigma^{\dagger} \Bigr\}_{\vec{k}},
\end{equation}
that sums over all permutations of an ordered Pauli string with $k_\sigma$ Pauli letters $\sigma \in \{X, Y, Z, I\}$.  We call these operators \emph{symmetrized Pauli strings}.
The identity $I^{\otimes n}$ is identified by $\vec{k}=(0,0,0,n)$. 
Since the basis is orthonormal, counting the distinct $\vec{k}$ satisfying the conditions enounced earlier gives us the dimension of $\piopspace$. Namely, $d_{\rm PI} = {\rm Te}(n+1)$, where ${\rm Te}(n):=n(n+1)(n+2)/2$ are the tetrahedral numbers. That is, the PI operators form a polynomial-dimension subspace of the $4^n$-dimension space of operators.

Often, we wish to express the symmetric part of an operator $A$ in the symmetrized Pauli basis. Assuming a decomposition $A=\sum_j w_j P_j$ in the Pauli basis ${P_j}$, we need to obtain overlaps between (non-symmetric) Pauli strings with the (symmetric) symmetrized ones. Given a Pauli string $P$ with composition $\vec{k}$, we have
\begin{equation}\label{eq:pauli_to_pauli_symm}
\trace[P B_{\vec{k'}}] = \delta_{\vec{k}, \vec{k}'} \sqrt{\frac{d k_X! k_Y! k_Z! k_I! }{n!}},
\end{equation}
that allows us to obtain the coefficients $\trace[AB_{\vec{k}}]$.

\subsection{The PI operator Schur basis}\label{app:symmetric_group_opschur}
An alternative basis of $\piopspace$ is found through the representation theory of $S_n$~\cite{harrow2005applications,bacon2006efficient}.
The basis of $\HH$ in which the action of $S_n$, defined in Eq.~\eqref{eq:app_act_sn}, block-decomposes is known as the \emph{Schur basis}. Let $\ket{\lambda, T_{\lm}, q_{\lm}}$ denote elements of the Scur basis. In such basis, for any $\sigma \in S_n$, its representation decompose as
\begin{equation}\label{eq:app_sn_in_schur}
    U_{\sigma} = \bigoplus_{\lambda} r_{\lambda}(\sigma) \otimes I_{m_\lambda}, 
\end{equation}
with $r_\lambda(\sigma) \in \mathbb{C}^{d_\lm \times d_\lm}$ the irreducible representations (irreps) that act on isomorphic subspaces $\mathcal{H}_\lm$ of dimension $d_\lm := {\rm dim}(\HC_{\lm})$, repeated $m_{\lm}$ times. 
The $\lambda$ label irreps of $S_n$ and are partition of the integer $n$ into at most $D$ parts ($D=2$ for qubits  and $D>2$ for general qudits). 
 
 Each partition can be associated to a Young diagram with $\lm_i$ boxes in its $i$-th row.
The $T_\lm$ labels elements of a basis of $\mathcal{H}_\lm$ and can be associated to Standard Young Tableaux: a filling of the Young diagram boxes with \emph{distinct} integers taking values in the set $[n]$ and arranged in strictly increasing order both row-wise and column-wise.
The $q_{\lm}$, labelling the multiplicities of the $S_n$-irrep, are in correspondence to Semi-Standard Young Tableaux: a filling of the Young diagram boxes with \emph{potentially repeated} integers taking values in the set $[D]$, and arranged in non-decreasing order row-wise and strictly increasing order column-wise. 
For qubits we can associate $q_{\lm}$ to half-integer values defined as follow. Given an irrep $\lm$, we define its (half-integer) spin value 
\begin{equation}
    s_\lm:= (\lm[0]-\lm[1])/2 \in [0, n/2].
\end{equation}
In turn, the valid $q_{\lm}$ can only take values in the set of half-integers
 \begin{equation}\label{eq:valid_q}
     Q_{\lm}:=\set{-s_\lm, -s_\lm+1, \hdots, s_\lm}.
 \end{equation}
Note that given that $\lm[0]\geq\lm[1]$ and $\lm[0]+\lm[1]=n$, at fixed $n$ we can recover $\lm$ from $s_{\lm}$ and sometimes we use the latter instead to indicate the former. For instance, we sometimes write $Q_{s_{\lm}}$ instead of $Q_\lm$.
Finally, we recall that the numbers of valid $T_{\lm}$ and $q_{\lm}$ corresponds to the dimensions $d_\lm$ and $m_\lm$ respectively.

As per Eq.~\eqref{eq:irrep_decomp_comm}, in the Schur basis, any PI operator $A$ has to be of the form
\begin{equation}\label{eq:app_irrep_decomp_comm}
    A \cong \bigoplus_{\lambda} I_{d_{\lambda}} \otimes A_{\lambda}, 
\end{equation}
with $A_{\lm} \in \mathbb{C}^{m_\lm \times m_{\lm}}$. Hence, an orthonormal basis of the PI operators space, that we call the \emph{PI operator Schur basis}, is given by the set of operators
\begin{equation}\label{eq:app:basis_rep_th}
    B_{\rm Schur} = \Bigr\{ B^{\lm}_{q_{\lm}, q'_\lm}:= \frac{1}{\sqrt{d_\lm}} \sum_{T_{\lm}} \ket{\lambda, T_{\lm}, q_{\lm}}\bra{\lambda, T_{\lm}, q'_{\lm}} \Bigr\}_{\lm, q_{\lm}, q'_{\lm}}.   
\end{equation}
We further note that
\begin{equation}
B^{\lm}_{q_{\lm}, q'_\lm} B^{\alpha}_{q_{\alpha}, q'_\alpha} = \frac{\delta_{\lm, \alpha} \delta_{q'_\lm, q_\alpha}}{d_{\lm}} B^{\lm}_{q_\lm, q'_{\alpha}}.
\end{equation}

\subsection{Change of basis and a few identities}\label{app:symmetric_group_change}
Details of the change of basis between the symmetrized Pauli basis of Eq.~\eqref{eq:app_basis_piop} and the PI operator Schur basis of Eq.~\eqref{eq:app:basis_rep_th} have been worked out in Ref.~\cite{anschuetz2022efficient} (Lemma13). Performing this change of basis boils down to evaluating the coefficients 
\begin{equation}\label{eq:app_schur_to_pauli}
    \trace \left[  B_{\vec{k}} B^{\lm}_{q_{\lm}, q'_\lm} \right] = \frac{1}{\sqrt{ d^\lm}} \sum_{T_{\lm}} \bra{\lambda, T_{\lm}, q'_{\lm}} B_{\vec{k}} \ket{\lambda, T_{\lm}, q_{\lm}}. 
\end{equation}
Given that the operators $B_{\vec{k}}$ are PI, they are of the form Eq.~\eqref{eq:app_irrep_decomp_comm}. Hence, each summand of Eq.~\eqref{eq:app_schur_to_pauli} is equal to another. Fixing a reference label $T^{0}_\lm$ for each $\lm$ we thus have:
\begin{equation}\label{eq:app_schur_to_pauli_0}
    \trace \left[ B_{\vec{k}} B^{\lm}_{q_{\lm}, q'_\lm}  \right] = \sqrt{ d^\lm} \bra{\lambda, T^0_{\lm}, q'_{\lm}} B_{\vec{k}} \ket{\lambda, T^0_{\lm}, q_{\lm}}. 
\end{equation}
For a convenient choice of $T^0_{\lm}$, Ref.~\cite{anschuetz2022efficient} shows that elements of the Schur basis can be expressed as the product
\begin{align}
\begin{split}
\ket{\lambda, T^0_{\lm}, q_{\lm}} &= \ket{\Psi}^{\otimes n -2 s_{\lm}} \ket{D^{s_{\lm}}_{s_{\lm}-q_{\lm}}}
\end{split}
\end{align}
of singlet states $\ket{\Psi}:= 1/\sqrt{2} \left( \ket{01} - \ket{10} \right)$ and Dicke states 
\begin{align}
\begin{split}
\ket{D^{n}_{k}}:= {n \choose k}^{-\frac{1}{2}} \sum_{\substack{x \in \set{0,1}^n \\ hw(x)=k} } \ket{x}.
\end{split}
\end{align}
This renders computations of the terms in Eq.~\eqref{eq:app_schur_to_pauli_0} relatively easy.

Here, we provide several relations of interest that will be used later on. These include decompositions of the generators of unitaries $W^{\otimes n}$ for $W \in \textsf{SU}(2)$ in the PI operator Schur basis:
\begin{align}\label{eq:app_schur_gen_sumx}
\begin{split}
    &\widetilde{X}:=\sum_{j=1}^n X_j = \sum_{\lm, q_{\lm}} \sqrt{d_\lm} \left( \alpha^{+}(\lm, q_{\lm})  B^{\lm}_{q_{\lm}+1, q_\lm} + \alpha^{-}(\lm, q_{\lm}) B^{\lm}_{q_{\lm}-1, q_\lm}  \right)\\
    &\widetilde{Y}:=\sum_{j=1}^n Y_j = \sum_{\lm, q_{\lm}} i\sqrt{f^\lm} \left( \alpha^{+}(\lm, q_{\lm})  B^{\lm}_{q_{\lm}+1, q_\lm} - \alpha^{-}(\lm, q_{\lm}) B^{\lm}_{q_{\lm}-1, q_\lm}  \right)\\
    &\widetilde{Z}:=\sum_{j=1}^n Z_j = \sum_{\lm, q_{\lm}}  2q_{\lm}\sqrt{f^\lm}  B^{\lm}_{q_{\lm}, q_\lm},
\end{split}
\end{align}
defined in terms of the coefficients:
\begin{equation}
      \alpha^{-}(\lm, q_{\lm}) := \sqrt{(s_\lm + q_\lm) (s_\lm - q_\lm + 1)}, \quad \text{ and} \;\;
      \alpha^{+}(\lm, q_{\lm}) := \sqrt{(s_\lm + q_\lm + 1) (s_\lm - q_\lm )}.
\end{equation}
Note that sums over $\lm$ and $q_\lm$, as appearing in Eq.~\eqref{eq:app_schur_gen_sumx}, are understood as first summing over $\lm$, followed by a sum over all the valid $q_\lm$ per $\lm$. That is, the second sum runs over $q_\lm \in Q_\lm$ as per Eq.~\eqref{eq:valid_q}. 

It follows from Eq.~\eqref{eq:app_schur_gen_sumx}, that the adjoint action (i.e., the action through commutation $[A, X]:=A X - X A$), of the generators ($\widetilde{X}$, $\widetilde{Y}$ and $\widetilde{Z}$) on PI operators is entirely captured by the following expressions:
\begin{align}\label{eq:app_schur_gen_sumx_adj}
\begin{split}
    [\widetilde{X}, B^{\lm}_{q_{\lm}, q'_{\lm}}] &=  \alpha^{+}(\lm, q_{\lm}) B^{\lm}_{q_{\lm}+1, q'_\lm}+ \alpha^{-}(\lm, q_{\lm})  B^{\lm}_{q_{\lm}-1, q'_\lm}  - \alpha^{+}(\lm, q'_{\lm}-1) B^{\lm}_{q_{\lm}, q'_\lm-1} \alpha^{-}(\lm, q'_{\lm}+1)  B^{\lm}_{q_{\lm}, q'_\lm +1} , \\
    [\widetilde{Y}, B^{\lm}_{q_{\lm}, q'_{\lm}}] &= i \left( \alpha^{+}(\lm, q_{\lm}) B^{\lm}_{q_{\lm}+1, q'_\lm} - \alpha^{-}(\lm, q_{\lm})  B^{\lm}_{q_{\lm}-1, q'_\lm} - \alpha^{+}(\lm, q'_{\lm}-1) B^{\lm}_{q_{\lm}, q'_\lm-1} + \alpha^{-}(\lm, q'_{\lm}+1)  B^{\lm}_{q_{\lm}, q'_\lm +1}  \right),\\
    [\widetilde{Z}, B^{\lm}_{q_{\lm}, q'_{\lm}}] &= 2\left(q_{\lm} - q'_{\lm} \right) B^{\lm}_{q_{\lm}, q'_\lm}.
\end{split}
\end{align} 

\subsection{Symmetrized PI-CS protocol}\label{app:bck_symm_pics}
Recall that the symmetrized PI-CS (dubbed \emph{symm-PI}) of Sec.~\ref{sec:pi_protocol_2} are obtained by (i) applying $W^{\otimes n}$ with a random single-qubit unitary $W$ and (ii) performing Hamming-weight projective measurements. We detail further these two steps.

The first step consists in sampling a random $W \in \SU(2)$.
This is achieved by parametrizing $W$ in terms of the Euler angles $\theta_1 \in [0, 2\pi]$, $\theta_2 \in [0, \pi]$ and $\theta_3 \in [0, 2\pi]$ with suitable probability distribution. Explicitly, taking
\begin{equation}\label{eq:param_su2}
W_{\thv}=e^{i \frac{\theta_3}{2} Z} e^{i \frac{\theta_2}{2} Y} e^{i \frac{\theta_1}{2} Z}\quad \text{with angles sampled s.t.} \;\;
\begin{cases}
    p(\theta_1) = p(\theta_3) = \frac{1}{2 \pi}\\
    p(\theta_2) = \frac{\sin(\theta_2)}{2}
\end{cases},
\end{equation}
ensures that we sample from the uniform (Haar) distribution.

After applying $W_{\thv}^{\otimes n}$, we perform a measurement $\set{\pihw}^n_{\hw=1}$ with the Hamming-weight projector $\pihw$ defined as 
\begin{align}\label{eq:rec_piproj}
\begin{split}
\pihw := \sum_{ \substack{x \in \{0,1\}^n\\ |x|=\hw} } \ket{b}\bra{b} &= \frac{1}{\hw ! (n-\hw)!} \sum_{\sigma \in S_n} U_\sigma \big( \ket{ \underbrace{0 \hdots 0}_{\hw} \underbrace{1 \hdots 1}_{n-\hw}} \bra{ \underbrace{0 \hdots 0}_{\hw} \underbrace{1 \hdots 1}_{n-\hw}} \big) U_\sigma^{\dagger} \\
&= \frac{1}{\hw ! (n-\hw)!} \sum_{\pi \in S_n} \pi \big( \ket{0}\bra{0}^{\otimes \hw} \ket{1}\bra{1}^{\otimes n-\hw} \big) \pi^{\dagger}
\end{split}
\end{align}
with $|x|$ being the Hamming-weight of the bitstring $x$. Here, we take the convention that it corresponds to the number of $0$s (rather than $1$s) contained in $x$. In the second line, the projector has been decomposed as a sum over all possible permutations of a projector onto an ordered bitstring (with all the $0$s first). The additional scaling factor accounts for the  fact that $\hw ! (n-\hw)!$ permutations leave the ordered projector unchanged. 

\section{Symmetrized PI-CS in the symmetrized Pauli basis}\label{app:mc_var_symm_pics}
In this section, we provide sufficient details for the implementation and use of the symmetrized PI-CS described in Sec.~\ref{app:bck_symm_pics} and based on an analysis performed in the symmetrized Pauli basis that was defined in Eq.~\eqref{eq:app_basis_piop}.
In Sec.~\ref{app:eval_mc_symm_pauli}, we derive expressions for the weights of the measurement channel.
In Sec.~\ref{app:const_cs_estimates}, we detail the necessary steps for the evaluation of CS estimates for expectation values. 
In Sec.~\ref{app:ev_cs_variances}, we present expressions for the resulting variances. Finally, in Sec.~\ref{app:qudits_paulibasis} we discuss extension to the qudit setting.

\subsection{Evaluating the measurement channel}\label{app:eval_mc_symm_pauli}
Here, we seek to evaluate the coefficients $c(\vec{k}, \vec{k}')$ appearing in Eq.~\eqref{eq:twirl_pitilde} in the symmetrized Pauli basis $B_{\rm Pauli}$. 
To do so, we will first express $\widetilde{\Pi}$ in $B_{\rm Pauli}$. Then, we will evaluate the $2n$-fold twirl $\E_{W} \big[ (W^{\dagger})^{\otimes 2n} \, \widetilde{\Pi} \, W^{\otimes 2n}\big]$ over the group of single-qubit unitaries $W \in \SU(2)$. 

\subsubsection{Projectors in the symmetrized Pauli basis} 
Given that the Hamming-weight projectors $\pihw$, defined in Eq.~\eqref{eq:rec_piproj}, are sums of projectors onto computational basis elements, they can only overlap with those symmetrized Pauli strings of Eq.~\eqref{eq:app_basis_piop} consisting of only $Z$ and $I$ Pauli operators, that is, with composition $\vec{k}$ of the form $(0,0,m, n-m)$. To simplify notations, we define 
\begin{equation}\label{eq:app_bm}
    B^m:=B_{(0,0,m, n-m)},
\end{equation}
with $m \in \{0, \hdots, n\}$. With these we can rewrite the Hamming-weight projectors as
\begin{equation}\label{eq:dec_pihw}
    \pihw = \sum_{m=0}^n \alpha(\hw, m) B^{m}.
\end{equation}
Given that each $B_m$ has norm one, we can evaluate the weights $\alpha(\hw, m)$ as
\begin{align}\label{eq:alpha_wm}
\begin{split}
\alpha(\hw, m) &:= \trace \bigg[ B^{m} \pihw \bigg] 
    = d^{-\frac{1}{2}} {n \choose m}^{\frac{1}{2}} a(\hw, m), \;\;\text{with} \;\;\; a(\hw, m) := \sum_{l=0}^{w} {m \choose l} {n-m \choose w-l} (-1)^{m-l}.
\end{split}
\end{align}
The coefficients $a(\hw, m)$ are sums of the expectation values of the operator $Z^{\otimes m} I^{\otimes n-m}$ for all distinct bitstrings with a number $\hw$ of $0$s and a number $n-\hw$ of $1$s. 

Recalling the definition of the $2n$-qubit operator $\widetilde{\Pi}$ of Eq.~\eqref{eq:app_mc}, we get that:
\begin{align}\label{eq:decomp_pitilde}
\begin{split}
    \widetilde{\Pi} &:= \sum_{w=0}^n (\pihw)^{\otimes 2} = \sum_{m,m'=0}^n \tilde{\alpha}(m,m') B^{m} \otimes B^{m'}, \quad \text{ with  }\;\; \tilde{\alpha}(m,m')=\sum_w \alpha(w, m) \alpha(w, m').
\end{split}
\end{align}
We note that the contribution of an operator $B^{m} \otimes B^{m'}$  to the sum is non-zero only when the absolute difference $|m-m'|$ is even. In such case, and defining $\Delta^{\pm}(m, m'):=|m \pm m'|$, we obtain:
\begin{align}
\begin{split}
    \alpha(m, m') = \alpha(\Delta^{\pm}) 
    &= 
\frac{(-1)^\frac{\Delta^-}{2}}{d} {n \choose m}^{\frac{1}{2}} {n \choose m'}^{\frac{1}{2}} \frac{\Delta^+! (2n - \Delta^+)!}{\frac{ \Delta^+}{2}! (n - \frac{\Delta^+}{2})! n!}.
\end{split}
\end{align}

\subsubsection{ 2$n$-fold averages.} We now compute averages of operators under the action of the $2n$-fold tensor product $W^{\otimes 2n}:$ 
\begin{equation}\label{eq:eu_conj_bmmprime}
     \overline{B^m \otimes B^{m'}}:= \E_{W} \big[ (W^{\dagger})^{\otimes 2n}  \left(B^{m} \otimes B^{m'}\right)  W^{\otimes 2n}\big],
\end{equation}
for arbitrary pairs of $(m,m')\in \{0, n\}^2$.
From the definitions of the operators $B^m$ in Eq.~\eqref{eq:app_bm} and~\eqref{eq:app_basis_piop}, we get that
\begin{align}
\begin{split}
B^{m} \otimes B^{m'} =
\frac{\sqrt{{n \choose m} {n \choose m'}}}{ (n!)^2 d} 
\sum_{\sigma, \sigma' \in Sn} \left(U_\sigma \otimes U_{\sigma'}\right) \left( Z^{\otimes m} I^{\otimes n-m} Z^{\otimes m'} I^{\otimes n-m'}  \right)   \left(U_\sigma \otimes U_{\sigma'}\right)^{\dagger}.
\end{split}
\end{align}
Given that any of the $W^{\otimes 2n}$ commutes with any of the $U_{\sigma} \otimes U_{\sigma'}$ it follows that
\begin{align}\label{eq:conj_bmmprime}
\begin{split}
\overline{B^m \otimes B^{m'}} = 
\frac{\sqrt{{n \choose m} {n \choose m'}}}{ (n!)^2 d} 
\sum_{\sigma, \sigma' \in Sn} \left(U_\sigma \otimes U_{\sigma'}\right)
\E_{W}  \bigg[
Z_W^{\otimes m} I^{\otimes n-m} Z_W^{\otimes m'} I^{\otimes n-m'}    
\bigg]
\left(U_\sigma \otimes U_{\sigma'}\right)^{\dagger}
\end{split}
\end{align}
where we have defined the single-qubit operator $Z_W := W Z W^{\dagger}$. 

When performing the integration appearing in Eq.~\eqref{eq:conj_bmmprime}, 
we can restrict our attention to the $M=m+m'$ qubits on which the $Z_W$ operators act, but leave aside the trivial contribution on the $2n-M$ qubits on which act the identities.
Following the treatment of Ref.~\cite{van2022hardware} (Eq.~B18 through B25), we know that
\begin{equation}\label{eq:exp_zum}
     \E_{W} \big[ (W^{\dagger})^{\otimes M} \, Z_W^{\otimes M} W^{\otimes M}\big] = \sum_{ \substack{M_X, M_Y, M_Z \in \mathbb{N} \\ M_X+M_Y+M_Z=M} }  v(M_X, M_Y, M_Z) \Bigg[ \sum_{\pi \in S_M} U_\pi \Big(X^{\otimes M_X} Y^{\otimes M_Y} Z^{\otimes M_Z} \Big) U_\pi^{\dagger} \Bigg].
\end{equation}
The coefficients $v(M_X, M_Y, M_Z)$ appearing in Eq.~\eqref{eq:exp_zum} depend on the numbers $M_{\alpha}$ of Paulis $\alpha \in \{X,Y,Z\}$ composing the ordered $M$-qubit Pauli string, and are defined as
\begin{equation}\label{eq:v}
     v(M_X, M_Y, M_Z) := \begin{cases}
    2 [\frac{M_X}{2}! \frac{M_Y}{2}! \frac{M_Z}{2}!]^{-1} \frac{(M/2 +1)!}{(M+2)!} & \text{for } M_X,M_Y,M_Z \in 2 \mathbb{N}\\
    0 & \text{otherwise}.
  \end{cases} 
\end{equation}

Plugging Eqs.~\eqref{eq:conj_bmmprime} and~\eqref{eq:exp_zum} into Eq.~\eqref{eq:eu_conj_bmmprime}  we get that
\begin{align}\label{eq:conj_bmmprime_full}
\begin{split}
&\overline{B^m \otimes B^{m'}} = \frac{\sqrt{{n \choose m} {n \choose m'}}}{ (n!)^2 d} \times
  \\
& \quad \sum_{\sigma, \sigma' \in Sn} \left(U_\sigma \otimes U_{\sigma'} \right) \Biggl( \sum_{ \substack{M_X, M_Y, M_Z \in \mathbb{Z} \\ M_X+M_Y+M_Z=M} }  v(M_X, M_Y, M_Z) \Bigg[ \sum_{\pi \in S_M} U_\pi \Big(X^{\otimes M_X} Y^{\otimes M_Y} Z^{\otimes M_Z} \Big) (U_\pi)^{\dagger} \Bigg] \Biggl)  \left(U_\sigma \otimes U_{\sigma'} \right)^{\dagger}.
\end{split}
\end{align}
Given that this operator is $S_n \times S_n$ invariant (i.e., is invariant under the action of independent permutations $\sigma$ and $\sigma'$ acting on the first and last $n$ qubits respectively), it can be expressed as 
\begin{equation}\label{eq:app_bmbmp_twirl}
\overline{B^m \otimes B^{m'}} = \sum_{\vec{k}, \vec{k'}} w(m, m', \vec{k}, \vec{k'}) \piop[k] \otimes \piop[k']
\end{equation}
in terms of the $d_{\rm PI}^2 = [{\rm Te}(n+1)]^2$ orthonormal operators 
\begin{align}\label{eq:AkAkp}
    \begin{split}
    & \piop[k] \otimes \piop[k'] = \frac{1}{n! d \sqrt{k_X! k_Y! k_Z! k_I! k'_X! k'_Y! k'_Z! k'_I!}} \times \\
    & \quad \quad \quad \quad \quad \sum_{\sigma, \sigma' \in Sn}  \left(U_\sigma \otimes U_{\sigma'}\right) \bigg( X^{\otimes k_X} Y^{\otimes k_Y} Z^{\otimes k_Z} I^{\otimes k_I}X^{\otimes k^{'}_X} Y^{\otimes k^{'}_Y} Z^{\otimes k^{'}_Z} I^{\otimes k^{'}_I} \bigg) \left(U_\sigma \otimes U_{\sigma'}\right)^{\dagger}.
\end{split}
\end{align}
The weights $w(m, m', \vec{k}, \vec{k'}):= \trace [\overline{B^m \otimes B^{m'}} \piop[k] \otimes \piop[k']]$ of Eq.~\eqref{eq:app_bmbmp_twirl} can only be non-zero when the number of identities in the summands of Eqs.~\eqref{eq:conj_bmmprime} and~\eqref{eq:AkAkp} agree on each $n$-qubit subsystem. That is, when $n-m=k_I$ and $n-m'=k'_I$. Similarly, summands in Eqs.~\eqref{eq:conj_bmmprime_full} and~\eqref{eq:AkAkp} only have non-zero overlap when the composition of the Paulis $\sigma\in\{X, Y, Z\}$ involved match ($M_{\sigma}=k_{\sigma}+k'_{\sigma}$). In such case, it remains to count how many permutations (out of the $(n!)^4 M!$ permutations involved) result in non-zero overlap (each of them evaluating to $\trace[I^{\otimes 2n}]=d^2$). Accounting for the various normalization factors and the conditions for which the weights $v(M_X, M_Y, M_Z)$ in Eq.~\eqref{eq:v} are non-zero, we get
\begin{align}\label{eq:wmmpkkp}
\begin{split}
&w(m, m', \vec{k}, \vec{k'}) = 2\frac{(\frac{m+m'}{2}+1)!}{ (m+m'+2)! }.
\frac{(k_X+k'_X)!}{(\frac{k_X+k'_X}{2})!}.
\frac{(k_Y+k'_Y)!}{(\frac{k_Y+k'_Y}{2})!}.
\frac{(k_Z+k'_Z)!}{(\frac{k_Z+k'_Z}{2})!}.
\sqrt{
\frac{m!m'!}{k_X!k'_X!k_Y!k'_Y!k_Z!k'_Z!},
}
\\
&\quad\quad\text{whenever} \begin{cases}
\text{(i)}\;\;k_I =n-m, \, k'_I =n-m', \\
\text{(ii)}\;k_X + k'_X \in 2\mathbb{N}, \, k_Y + k'_Y \in 2\mathbb{N}, \,k_Z + k'_Z \in 2\mathbb{N}, 
\end{cases}, \quad \text{and otherwise} \; 0.
\end{split}
\end{align}

\subsubsection{Measurement channel and its inverse} Finally, aggregating all contributions through Eq.~\eqref{eq:decomp_pitilde}, we obtain the sought-after weights
\begin{align}\label{eq:res_w_k_kp}
\begin{split}
&c(\vec{k}, \vec{k'}) = \sum^n_{m,m'=0} \alpha(m,m') w(m, m', \vec{k}, \vec{k'}) = \alpha(n-k_I,n-k'_I) w(n-k_I,n-k'_I, \vec{k}, \vec{k'})\\
&\quad \quad \quad \;  = 
\frac{(-1)^\frac{|k_I-k'_I|}{2}}{d(2n-k_I-k'_I+1)
\sqrt{
k_X!k'_X!k_Y!k'_Y!k_Z!k'_Z!k_I!k'_I!
}}.
\frac{(k_X+k'_X)!}{(\frac{k_X+k'_X}{2})!}.
\frac{(k_Y+k'_Y)!}{(\frac{k_Y+k'_Y}{2})!}.
\frac{(k_Z+k'_Z)!}{(\frac{k_Z+k'_Z}{2})!}
\frac{(k_I+k'_I)!}{(\frac{k_I+k'_I}{2})!}
 \\
& \quad \text{whenever} \begin{cases}
\text{(i)}\;\;\;k_X + k'_X \in 2\mathbb{N} \\
\text{(ii)}\; k_Y + k'_Y \in 2\mathbb{N} \\
\text{(iii)}\, k_Z + k'_Z \in 2\mathbb{N} 
\end{cases}, \quad \text{and otherwise} \; 0.
\end{split}
\end{align}

The weights of Eq.~\eqref{eq:res_w_k_kp} entirely characterize the measurement channel $\mathcal{M}_{\rm symm-PI}$ of the symmetrized PI-CS, as per Eq.~\eqref{eq:mc_symm}, and can be organized as a $d_{\rm PI} \times d_{\rm PI}$ matrix $\W$ such that $\W_{\vec{k}, \vec{k'}}=c(\vec{k}, \vec{k'})$. 
The cost of inverting the channel is dominated by the inversion of $\W$. It can be reduced by choosing an ordering of the vectors $\vec{k}=(k_X, k_Y, k_Z, k_I)$ to block diagonalizes $\W$. Given the conditions for the weights to be non-zero in Eq.~\eqref{eq:res_w_k_kp}, the vectors $\vec{k}$ can be ordered in terms of the parities of its elements $k_X$, $k_Y$ and $k_Z$, yielding a total of $8$ blocks for any value of $n$.

\subsubsection{Bounds on the variance}\label{app:var_o_symmcs}
Numerical eigendiagonalization of $\W$ reveals that its smallest eigenvalue is given by $\lm_0= 1/(2n+1)$. Direct application of Eq.~\eqref{eq:app_var_o} yields the bounds on the variance:
\begin{align}\label{eq:app_var_o_symmc}
\begin{split}
     \var[\hat{o}] \leq (2n+1) ||O||^2.
\end{split}
\end{align}
This tells us, for instance, that when estimating fidelities (i.e., expectation value of a projector), the variance in the estimates obtained through the symmetrized CS can grow at most linearly with the system size.

\subsection{Constructing the classical shadow estimates}\label{app:const_cs_estimates}
As per Eq.~\eqref{eq:app_cs_est_exp}, to construct the CS estimates of the expectation values $o=\trace[O \rho]$, we need to evaluate
\begin{equation}
    \trace \bigg[O \mathcal{M}_{\rm symm}^{-1}\Big( (W_{\thv}^{\dag})^{\otimes n} \pihw  W_{\thv}^{\otimes n}\Big) \bigg]
\end{equation}
for the random single-qubit unitaries $W_{\thv}$ that have been sampled through the parametrization of Eq.~\eqref{eq:param_su2} and for the Hamming-weight projectors $\pihw$ of Eq.~\eqref{eq:rec_piproj}. 
Given that we have obtained the action  of the measurement channel (and hence its inverse) in the basis $B_{\rm Pauli}=\set{B_{\vec{k}}}$, we keep performing our analysis in such basis. 
In particular, we now  seek to express the operators $(W^{\dag})^{\otimes n} \pihw W^{\otimes n}$, for arbitrary $U$ and $\hw$, in terms of $B_{\vec{k}}$. 

First, defining the conjugated operator $Z_{\thv}:=W^{\dagger}_{\thv} Z W_{\thv}$ and recalling that any $W_{\thv}^{\otimes n}$ commute with any $U_{\sigma}$ we get:
\begin{align}\label{eq:app_tmp}
\begin{split}
    (W_{\thv}^{\dag})^{\otimes n} B^m  W_{\thv}^{\otimes n} 
    &= \frac{1}{\sqrt{n!m!(n-m)!d}}  \sum_{\sigma' \in S_n} U_{\sigma'} (Z_{\thv}^{\otimes m}I^{\otimes n-m})  U_{\sigma'}^{\dag}, 
\end{split}
\end{align}
for the operators $B^m$ defined in Eq.~\eqref{eq:app_bm}.
Next, given 
the parametrization $W_{\thv}=e^{i \frac{\theta_3}{2} Z} e^{i \frac{\theta_2}{2} Y} e^{i \frac{\theta_1}{2} Z}$ in  Eq.~\eqref{eq:param_su2}, we obtain the decomposition
$Z_{\thv} = Z_{\thv, X} X + Z_{\thv, Y} Y + Z_{\thv, Z} Z$ with coefficients evaluating to 
\begin{align}
\begin{split}    
       Z_{\thv, X} = \sin(\theta_2) \cos(\theta_1),\quad
       Z_{\thv, Y} = \sin(\theta_2) \sin(\theta_1),\; \quad \text{and }\;
       Z_{\thv, Z} = \cos(\theta_2).
\end{split}
\end{align}
We can now compute the overlaps between the operators in Eq.~\eqref{eq:app_tmp} and the symmetrized Paulis $B_{\vec{k}}$ in Eq.~\eqref{eq:app_basis_piop}. These overlaps are non-zero only if the number of identities involved in each agree, i.e., if $n-m=k_I$. In such case, 
\begin{align}
\begin{split}
    \trace [ &\piop   (W_{\thv}^{\dag})^{\otimes n} B^m  W_{\thv}^{\otimes n}] 
    \\ &= \frac{1}{n! k_I! d \sqrt{(n-k_I)!k_X!k_Y!k_Z!}} \trace \bigg[ \left(\sum_{\sigma \in S_n} U_\sigma (X^{\otimes k_X}Y^{\otimes k_Y}Z^{\otimes k_Z}I^{\otimes k_I}) U_\sigma^{\dag}\right) \left( \sum_{\sigma' \in S_n} U_{\sigma'} (Z_{\thv}^{\otimes m}I^{\otimes n-m})  U_{\sigma'}^{\dag}\right) \bigg] \\
    &= \frac{1}{d k_I! \sqrt{(n-k_I)! k_X! k_Y! k_Z! } } \trace \bigg[ \Big(X^{\otimes k_X}Y^{\otimes k_Y}Z^{\otimes k_Z}I^{\otimes k_I} \Big) \Big(\sum_{\pi' \in S_n} \pi' (Z_U^{\otimes (n-k_I)}I^{\otimes k_I}) \pi'^{\dag}\Big) \bigg] \\
    &=  \frac{ \sqrt{(n-k_I)!}}{\sqrt{k_X! k_Y! k_Z! } } (Z_{\thv, X})^{k_X} (Z_{\thv, Y})^{k_Y} (Z_{\thv, Z})^{k_Z}
\end{split}
\end{align}
Finally, recalling the decomposition of $\pihw$ from Eq.~\eqref{eq:dec_pihw} in terms of the weights of Eq.~\eqref{eq:alpha_wm}, we get that
\begin{align}
\begin{split}
    \trace [ \piop   (W_{\thv}^{\dag})^{\otimes n} \pihw  W_{\thv}^{\otimes n}] 
    &= \sum_{m=0}^n \alpha(\hw, n-k_I) \trace [ (W^{\dag})^{\otimes n} \piop B^m  W^{\otimes n} ] \\
    &=   \frac{\sqrt{n!}}{\sqrt{k_I! k_X! k_Y! k_Z! d} } a(\hw, n-k_I)  (Z_{\thv, X})^{k_X} (Z_{\thv, Y})^{k_Y} (Z_{\thv, Z})^{k_Z}.
\end{split}
\end{align}
This gives us the entries of the vector $\ket{(W_{\thv}^{\dag})^{\otimes n} \pihw  W_{\thv}^{\otimes n}} \rangle$ on which acts the inverse channel matrix $\W^{-1}$. Together with Eq.~\eqref{eq:pauli_to_pauli_symm}, that gives the components of any Pauli operator in the symmetrized Pauli basis, this allows us to evaluate the CS expectation values estimates as per Eq.~\eqref{eq:vec_cs_est}. 

\subsection{Evaluating the variances entailed}\label{app:ev_cs_variances}
We aim at evaluating the variance of the CS estimates defined in Eq.~\eqref{eq:app_var}, and seek to obtain the weights of Eq.~\eqref{eq:twirl_pi_triple} entering the expression of the variance of Eq.~\eqref{eq:var_symm}.
To do so, we follow a similar approach as detailed in Sec.~\ref{app:eval_mc_symm_pauli}, and first express $\doublewidetilde{\Pi}$ in the basis $B_{\rm Pauli}$, and then we evaluate the $3n$-fold twirl $\E_{W} \big[ (W^{\dagger})^{\otimes 3n} \, \doublewidetilde{\Pi} \, W^{\otimes 3n}\big]$.

First, one can rewrite the 3n-qubit operator $\doublewidetilde{\Pi}$, defined in Eq.~\eqref{eq:app_var} as 
\begin{align}\label{eq:pi_three}
\begin{split}
    \doublewidetilde{\Pi} &:= \sum_{\hw=0}^n (\pihw)^{\otimes 3}   = \sum_{m,m',m''=0}^n \alpha(m, m',m'') B^{m} \otimes B^{m'} \otimes B^{m''}
\end{split}
\end{align}
in terms of the 3-fold tensor of PI operators $B^{m} \otimes B^{m'} \otimes B^{m''}$, and that has coefficients
\begin{align}\label{eq:alpha_mmpmppp}
\begin{split}
    \alpha(m, m',m'') = \sum_{\hw=0}^n  \alpha(\hw, m) \alpha(\hw, m') \alpha(\hw, m''),
\end{split}
\end{align}
with $\alpha(\hw, m)$ previously defined in Eq.~\eqref{eq:alpha_wm}.
Then, we can compute the overlap between the twirled operators 
\begin{align}\label{eq:eu_conj_bmmpmpp}
\begin{split}
 &\overline{B^m \otimes B^{m^{'}}\otimes B^{m^{''}}}  := \E_{W} \big[ (W^{\dagger})^{\otimes 3n} \, B^{m} \otimes B^{m'} \otimes B^{m''} \, W^{\otimes 3n}\big] \\
&= \frac{\sqrt{{n \choose m} {n \choose m'} {n \choose m''} }}{ (n!)^3 d^{\frac{3}{2}}}\sum_{\sigma, \sigma', \sigma^{''} \in Sn} \left(U_\sigma \otimes U_{\sigma'}\otimes U_{\sigma^{''}}\right)
\E_{W}  \bigg[
Z_W^{\otimes m} I^{\otimes n-m} Z_W^{\otimes m'} I^{\otimes n-m'}  Z_U^{\otimes m''} I^{\otimes n-m''}    
\bigg]
\left(U_\sigma \otimes U_{\sigma'}\otimes U_{\sigma^{''}}\right)^{\dagger}
\end{split}
\end{align}
for arbitrary triplets of $(m,m',m'')\in \{0, \hdots, n\}^3$, and the $d_{\rm PI}^3 = [{\rm Te}(n+1)]^3$ orthonormal operators 
\begin{align}\label{eq:AkAkpAkpp}
    \begin{split}
    & \piop[k] \otimes \piop[k'] \otimes \piop[k''].
\end{split}
\end{align}
That is, we would evaluate the weights
\begin{equation}
    w(m,m',m'', \vec{k}, \vec{k'}, \vec{k''}):= \trace \bigg[ \Big( \overline{B^{\otimes m} \otimes B^{m'} \otimes B^{m''}} \Big) \Big( \piop[k] \otimes \piop[k'] \otimes \piop[k'] \Big) \bigg].
\end{equation}
As before, these weights are non-zero only when $m=n-k_I$, $m'=n-k'_I$, $m''=n-k''_I$, i.e., whenever the number of identities in both operators match on each of the $3$ n-qubit subsystems. Adopting the same reasoning as for Eq.~\eqref{eq:wmmpkkp}, we find
\begin{align}
\begin{split}
&w(m, m', m'', \vec{k}, \vec{k'}, \vec{k''}) 
\\&= 2\frac{(\frac{m+m'+m''}{2}+1)!}{ (m+m'+m''+2)! }.
\frac{(k_X+k'_X+k''_X)!}{(\frac{k_X+k'_X+k''_X}{2})!}.
\frac{(k_Y+k'_Y+k''_Y)!}{(\frac{k_Y+k'_Y+k''_Y}{2})!}.
\frac{(k_Z+k'_Z+k''_Z)!}{(\frac{k_Z+k'_Z+k''_Z}{2})!}.
\sqrt{
\frac{m!m'!m''!}{k_X!k'_X!k''_X!k_Y!k'_Y!k''_Y!k_Z!k'_Z!k''_Z!},
}
\\
&\text{whenever} \begin{cases}
\text{(i)}\;\;k_I =n-m, \, k'_I =n-m', k''_I =n-m''\\
\text{(ii)}\;(k_X + k'_X + k''_X) \in 2\mathbb{N}, \, (k_Y + k'_Y +k''_Y) \in 2\mathbb{N}, \,(k_Z +k'_Z + k''_Z)  \in 2\mathbb{N}, 
\end{cases} , \quad \text{and otherwise} \; 0.
\end{split}
\end{align}

Finally we get from Eq.~\eqref{eq:alpha_mmpmppp} that 
\begin{align}
\begin{split}
    \E_{W} \big[ (W^{\dagger})^{\otimes 3n} \, \doublewidetilde{\Pi} \, W^{\otimes 3n}\big] = \sum_{\vec{k}, \vec{k'}, \vec{k''}} c(\vec{k}, \vec{k'}, \vec{k''}) \piop[k] \otimes \piop[k'] \otimes \piop[k']
\end{split}
\end{align}
with the weights
\begin{align}\label{eq:res_w_k_kp_kpp}
\begin{split}
& c(\vec{k}, \vec{k'}, \vec{k''}) = \alpha(n-k_I,n-k'_I, n-k''_I) w(n-k_I,n-k'_I,n-k''_I, \vec{k}, \vec{k'},\vec{k''}) \\
& \text{whenever} \begin{cases}
\text{(i)}\;\;k_I =n-m, \, k'_I =n-m', k''_I =n-m''\\
\text{(ii)}\;(k_X + k'_X + k''_X) \in 2\mathbb{N}, \, (k_Y + k'_Y +k''_Y) \in 2\mathbb{N}, \,(k_Z +k'_Z + k''_Z)  \in 2\mathbb{N}, 
\end{cases} , \quad \text{and otherwise} \; 0.
\end{split}
\end{align}
These weights allow us to evaluate the variance in Eq.~\eqref{eq:app_var}.
\subsection{Qudits}\label{app:qudits_paulibasis}
Let us conclude by briefly discussing steps to be taken to port the expressions derived to qudit systems.
Let $D$ be the dimension of a single qudit, and denote as $\mathcal{H}^{(D)}$ the Hilbert space of the $n$-qudit system with dimension $d=D^n$.
The set $\set{\ket{j}}$ with $j \in \set{0, \hdots, D-1}^n$ labels the computational basis, and the set $\set{ Q_{l_1} \otimes \hdots \otimes Q_{l_n} }$ with $l \in [D^2]$ denotes a basis of $D^{2n}$ generalized Pauli strings. 

Akin to the qubit case in in Eq.~\eqref{eq:A_k}, a basis of the PI qudit operator space $\mathcal{L}(\mathcal{H}^{(D)})$ can be obtained through symmetrization of the generalized Pauli strings by twirling. This yields the orthonormal basis
\begin{equation}\label{eq:app_basis_piop_qudit}
    B^{(D)}_{\rm Pauli}=\Biggr\{ B_{\vec{t}} := \left( d n! \prod_{l=1}^n \vec{t}_l! \right)^{-\frac{1}{2}}\sum_{\sigma \in Sn}  U_\sigma \left( \bigotimes_{j=1}^{D^2} Q_j^{\otimes t_j} \right) U_\sigma^{\dagger} \Biggr\}_{\vec{t}},
\end{equation}
for any type $\vec{t} \in T(n, D^2)$ with entries $\vec{t}_j$. 
Recall from the main text that $T(n, s)$ is the set of all valid types for a string composed of $n$ symbols that can each take $s$ different values. That is, for $\vec{t}\in T(n, s)$, the type $\vec{t}\in \mathbb{N}^s$ satisfies $\sum \vec{t}_j = n$. Counting such valid types gives us the dimension of the PI operators:
\begin{equation}
{\rm dim}(\LL^{PI}(\mathcal{H}^{(D)})) = {n+D^2-1 \choose D^2-1}.    
\end{equation}

We further recall that the symmetrized PI-CSs for qudit systems consists in applying a unitary $W^{\otimes n}$, with $W$ drawn randomly from $\SU(D)$, followed by type measurement $\{\Pi^\TT_t \}_{\vec{t} \in T(n, D)}$ with projectors defined as 
\begin{equation}\label{eq:app_projectors_cb_qudits}
    \Pi_{\vec{t}}:=\sum_{\substack{x \in \{0,\hdots, D-1\}^{n}:\\ \vec{t}(x)=\vec{t}}} \ket{x}\bra{x}.
\end{equation}

As detailed in the main text, this symmetrized PI-CSs for qudit complies with the symmetric criterion of Eq.~\eqref{eq:symm_CS_rule}. We now sketch modifications required to obtain an expression of its measurement channel. 

Among the operators $B_{\vec{t}}$ of Eq.~\eqref{eq:app_basis_piop_qudit}, let us denote the $D^n$ ones diagonal in the computational basis as $B^{\vec{m}}$ with $\vec{m} \in T(n, D)$.
A first step is to derive overlaps between these operators and the projectors of Eq.~\eqref{eq:app_projectors_cb_qudits}. 
In turn, akin to Eq.~\eqref{eq:decomp_pitilde}, we would obtain weights of  
$\widetilde{\Pi} := \sum_{\vec{t}} (\Pi_{\vec{t}})^{\otimes 2}$ in terms of operators $B^{\vec{m}} \otimes B^{\vec{m}'}$.
Next, we would need to derive the coefficients $v(M_1, \hdots, M_{D^2 - 1})$ appearing in the decomposition
\begin{equation}\label{eq:exp_zum_qudit}
     \E_{W} \big[ (W^{\dagger})^{\otimes M} \, Z_W^{\otimes M} W^{\otimes M}\big] = \sum_{ \substack{M_l \in \mathbb{N} \\ \sum M_l = M} }  v(M_1, \hdots, M_{D^2-1}) \Bigg[ \sum_{\pi \in S_M} U_\pi \Big(\bigotimes_{l=1}^{D^2-1} Q_l^{\otimes M_l} \Big) U_\pi^{\dagger} \Bigg],
\end{equation}
where we recall that $Z_W := W^{\dagger} Z W$ with now $W \in \SU(D)$. These coefficients generalize Eq.~\eqref{eq:v} to the case of qudits and could be worked out following the treatment of Ref.~\cite{van2022hardware} (Eq.~B18 through B25) given a Euler angle parameterization of the unitaries $U\in \SU(d)$ and appropriate distributions over the angles~\cite{tilma2002generalized}. Finally, following derivations from Eq.~\eqref{eq:conj_bmmprime_full} to~\eqref{eq:res_w_k_kp_kpp}, one would obtain the desired weights.

\section{Symmetrized PI-CS in the Schur operator basis}\label{app:mc_schur_basis}

One can readily use the symmetrized PI-CS through the expressions in the symmetrized Pauli basis provided in the previous appendix. However, while maybe less natural, we found it beneficial to express and invert the measurement channel in an alternative basis. In Sec.~\ref{app:eval_mc_symm_schur}, we derive expressions for the measurement channel in the PI Schur operator basis, $B_{\rm Schur}$ defined in Eq.~\eqref{eq:app:basis_rep_th}. In Sec.~\ref{app:block_schur} we detail how, for this choice of basis, the matrix of the measurement channel decomposes in $2n+1$ blocks, thus easing its inversion.
Finally in Sec.~\ref{app:const_cs_estimates_schur}, we provide additional expressions to compute the CS estimates in this basis.

\subsection{Measurement channel in the Schur basis of PI operators}\label{app:eval_mc_symm_schur}
Our goal is to evaluate the coefficients appearing in the decomposition of Eq.~\eqref{eq:twirl_pitilde}: We seek the overlaps of 
\begin{equation}
\TT^{(2n)}[\widetilde{\Pi}] := \E_{W\sim \textsf{SU}(2)} [W^{\otimes 2n} \widetilde{\Pi} (W^{\otimes 2n})^{\dag}],
\end{equation}
with $\TT^{(2n)}$ being a twirl with respect to the $2n$-fold tensor products $W^{\otimes 2n}$, with the operators
\begin{equation}
    \tilde{A}^{\lm, \nu}_{(q_{\lm}, q'_\lm), (q_{\nu}, q'_\nu)}:= B^{\lm}_{q_{\lm}, q'_\lm} \otimes B^{\nu}_{q_{\nu}, q'_\nu},
\end{equation}
where the operators $B^{\lm}_{q_{\lm}, q'_\lm}$ have been defined in Eq.~\eqref{eq:app:basis_rep_th}. We follow the same approach as in Sec.~\ref{app:eval_mc_symm_pauli}.

The Hamming-weight projectors $\pihw$ admit a very compact form in $B_{\rm Schur}$. Defining $q(h):=w - \frac{n}{2}$ we have
\begin{equation}\label{eq:piw_piopsschur}
    \Pi_w = \sum_{\substack{\lm: q(h) \in Q_{\lm}} } B^{\lm}_{q(h), q(h)} \sqrt{d_\lm},
\end{equation}
with the sum indicating that only the irreps $\lm$ such that $q(h) \in Q_{\lm}$, with $Q_{\lm}$ defined in Eq.~\eqref{eq:valid_q}, contribute. It follows that
\begin{equation}\label{eq:def_pi_tilde}
    \widetilde{\Pi}:= \sum_h (\Pi_\hw)^{\otimes 2} = \sum_{q \in Q_n} \Big( \sum_{\substack{\lambda: q \in Q_{\lm} }} \sqrt{d_\lm} B^{\lm}_{(q, q)} \Big)^{\otimes 2} = \sum_{q \in Q_n} \sum_{\substack{\lambda: q \in Q_{\lm} \\ \alpha: q \in Q_{\alpha} }} \sqrt{d_\lm d_\alpha} \tilde{A}^{\lm, \alpha}_{(q, q), (q, q)}
\end{equation}
which is the alter-ego of Eq.~\eqref{eq:decomp_pitilde}.

To evaluate $\TT^{(2n)}[\widetilde{\Pi}]$, we start by recalling from Sec.~\ref{app:bck_symm} that the twirl $\TT^{(2n)}$ is an orthogonal projection onto the space of operators commuting with $W^{\otimes 2n}$. 
An orthogonal basis of such space can be written in terms of the $2n$-qubit Schur basis:
\begin{equation}\label{eq:basis_rep_th_u2_2n}
    B^{2n} = \Bigr\{ \widetilde{B}^{\mu}_{T_{\mu}, T'_\mu} := \frac{1}{\sqrt{m_\mu}} \sum_{q_{\mu}} \ket{\mu, T_{\mu}, q_{\mu}}\rangle \langle \bra{\mu, T'_{\mu}, q_{\mu}} \Bigr\}_{\mu, T_{\mu}, T'_{\mu}}.   
\end{equation}
We emphasize that, as opposed to the basis $B_{\rm Schur}$ in Eq.~\eqref{eq:app:basis_rep_th}, these operators act on $2n$ qubits (highlighted by the tilde for the operators, and the double kets for the vectors) and that now the indexing runs over the (exponentially many) pairs of $T_{\mu}$ and $T'_{\mu}$. We always use $\mu$ to refer to $2n$-qubit irreps. In such basis, we can write the twirl as
\begin{equation}\label{eq:twirl_2n}
    \TT^{(2n)}[\widetilde{\Pi}] = \sum_{T_\mu, T'_{\mu}} \trace [\widetilde{B}^{\mu}_{T_{\mu}, T'_\mu} \widetilde{\Pi}] \widetilde{B}^{\mu}_{T_{\mu}, T'_\mu}.
\end{equation}

To proceed further, we need to recall that the coupling of two $n$-qubit Schur basis states results in a superposition of $2n$-qubit Schur states
\begin{align}
\begin{split}
&\ket{\lambda, T_{\lm}, q_{\lm}} \ket{\alpha, T_{\alpha}, q_{\alpha}} = \sum_{\substack{\mu: \\ \mu \doteq \lambda + \alpha \\ q_\lm + q_\mu \in Q_{\mu}}} C^{q_\lm, q_{\alpha}}_{\lm, \alpha; \mu} \ket{ \mu, (T_{\lm}, T_{\alpha})^{\mu}, q_{\lm} + q_{\alpha}  } \rangle \\
&\textrm{with }\quad C_{\lm, \alpha; \mu}^{q_\lm, q_\alpha}:= CG(s_\lm, s_\alpha, s_\mu, q_\lm, q_\alpha, q_\lm + q_\alpha),
\end{split}
\end{align}
defined in terms of the Clebsch-Gordan coefficients $CG(j_1, j_2, J, m_1, m_2, M)$.
The latter is non-zero only if $M=m_1+m_2\in Q_J$ and provided that $s_{\mu} \in \{|s_{\lm}-s_{\alpha}|, \hdots , s_{\lm}+s_{\alpha}\}$.
For given irreps $\lm$ and $\alpha$, we denote as $\mu \doteq \lm + \alpha$ the resulting irreps that satisfy the latter condition. 
Standard tableaux $T_{\lm}$ and $T_{\alpha}$ can only couple to a single standard tableau denoted $(T_{\lm}+T_{\alpha})^{\mu}$ per resulting irrep $\mu\doteq \lm + \alpha$. Given irreps $\lm$, $\alpha$, $\mu$ we denote the set of all possible resulting tableaux $T^{\mu}(\alpha, \mu)$, that is empty whenever $\mu \not\doteq \lm + \alpha$. 
We note that the Clebsch-Gordan mentioned refer to the case of qubit systems (i.e., refer to the representation theory of $\SU(2)$) which is our main focus. Still, they can be generalized to qudits~\cite{alex2011numerical} (i.e., for the representation theory of $\SU(D)$).

\medskip
\noindent

Given $T_{\mu}=(T_\lm + T_\alpha)^{\mu}$, we are now in position to evaluate
\begin{align}
\begin{split}\label{eq:trace_one}
    \trace [\widetilde{B}^{\mu}_{T_{\mu}, T'_\mu} \tilde{A}^{\nu, \beta}_{(q_{\nu}, q'_\nu), (q_{\beta}, q'_\beta)} ] 
    &= \frac{1}{\sqrt{d_{\nu} d_{\beta} m_{\mu}}} \sum_{q_{\mu},T_\nu, T_{\beta}} 
    \Big( \bra{\nu, T_{\nu}, q'_{\nu}} \bra{\beta, T_{\beta}, q'_{\beta}}  \ket{\mu, T_{\mu}, q_{\mu}}\rangle \Big) 
    \Big(\langle \bra{\mu, T'_{\mu}, q_{\mu}} \ket{\nu, T_{\nu}, q_{\nu}} \ket{\beta, T_{\beta}, q_{\beta}} \Big) \\
    &= \mathbbm{1}_{[T_\mu = T'_{\mu}]} \mathbbm{1}_{[\nu = \lm]} \mathbbm{1}_{[\beta = \alpha]} \mathbbm{1}_{[q_\nu + q_\beta = q'_\nu + q'_\beta]} \mathbbm{1}_{[q_\nu + q_\beta \in Q_{\mu}]} \frac{ C_{\nu, \beta; \mu}^{q'_\nu, q'_\beta}  C_{\nu, \beta; \mu}^{q_\nu, q_\beta}}{\sqrt{d_{\nu} d_{\beta} m_{\mu}}},
\end{split}
\end{align}
with the functions $\mathbbm{1}_{[{\rm C}]} =1$ if the condition ${\rm C}$ is satisfied and $0$ otherwise.
These conditions arise from orthonormality of the Schur basis vectors and Clebsch-Gordan rules discussed earlier.
We can specialize Eq.~\eqref{eq:trace_one} to: 
\begin{align}
\begin{split}\label{eq:trace_two}
    \trace [\widetilde{B}^{\mu}_{T_{\mu}, T_\mu} \tilde{A}^{\lm, \alpha}_{(q, q), (q, q)} ] 
    &= \mathbbm{1}_{[2q \in Q_{\mu}]} \frac{ \big(C_{\lm, \alpha; \mu}^{q, q}\big)^2}{\sqrt{d_{\lm} d_{\alpha} d_{\mu}}} \quad \quad \Rightarrow \quad \quad 
    \trace [\widetilde{B}^{\mu}_{T_{\mu}, T_\mu} \widetilde{\Pi}] 
    =  \sum_{\substack{q : \\ q \in Q_\lm \cap Q_{\alpha}\\2q \in Q_{\mu}}}  \frac{\big(C_{\lm, \alpha; \mu}^{q, q}\big)^2}{\sqrt{d_{\mu}}}=: \frac{\widetilde{\Pi}_{\mu, \lambda, \alpha}}{\sqrt{d_{\mu}}}.
\end{split}
\end{align}
Finally we can evaluate Eq.~\eqref{eq:twirl_2n} as
\begin{align}
\begin{split}
    \TT^{(2n)}[\widetilde{\Pi}] &= \sum_{\lm, \alpha} \sum_{\mu \doteq \lm + \alpha} \frac{\widetilde{\Pi}_{\mu, \lm, \alpha} \sqrt{d_{\lm} d_{\alpha}}}{m_{\mu}}  \left( \sum_{\substack{q_\lm ,q_\alpha , q'_\lm , q'_\alpha: \\ q_\lm + q_\alpha = q'_\lm + q'_\alpha \in Q_{\mu}}}  C_{\lm, \alpha; \mu}^{q'_\lm, q'_\alpha}  C_{\lm, \alpha; \mu}^{q_\lm, q_\alpha} \tilde{A}^{\lm, \alpha}_{(q_{\lm}, q'_\lm), (q_{\alpha}, q'_\alpha)} \right).
\end{split}
\end{align}
We have used the fact that only operators with $T'_{\mu}=T_\mu$ contribute, and rewrote the summations over all the $T_{\mu}$ as a summation first over all underlying $\lambda$ and $\alpha$, then over the valid $\mu$. Given that none of the quantities involved in the sum depend on explicit details of such tableaux they yield a constant multiplicative factor $|T^{\mu}(\lm, \alpha)|=d_{\lm}d_{\alpha}$.
It follows that, as per Eq.~\eqref{eq:mc_symm}, the measurement channel is fully specified as
\begin{align}\label{eq:mc_symm_schur}
\begin{split}
     \MM_{\rm symm}(\rho) 
     &= \sum_{k, k'} c(\lm, q_{\lm}, q'_\lm, \alpha, q_\alpha, q'_{\alpha}) \trace[\rho B^{\lm}_{q_{\lm}, q'_\lm}] B^{\alpha}_{q_{\alpha}, q'_\alpha},
\end{split}
\end{align}
in terms of the coefficients
\begin{align}
\begin{split}\label{eq:coeffs_mc_schur}
    c(\lm, q_{\lm}, q'_\lm, \alpha, q_\alpha, q'_{\alpha}) &:= \mathbbm{1}_{[q_\lm + q_\alpha =  q'_{\lm}+q'_{\alpha}]}  \sum_{\substack{\mu: \\ \mu \doteq \lm + \alpha \\  q_\lm + q_\alpha \in Q_{\mu}} } \frac{\widetilde{\Pi}_{\mu, \lm, \alpha} \sqrt{d_{\lm} d_{\alpha}}}{m_{\mu}}  \left(   C_{\lm, \alpha; \mu}^{q'_\lm, q'_\alpha}  C_{\lm, \alpha; \mu}^{q_\lm, q_\alpha} \right).
\end{split}
\end{align}

\subsection{Block diagonalization}\label{app:block_schur}
We now verify that writing the measurement channel in $B_{\rm Schur}$ is indeed beneficial. To see this, let us organise the coefficients in Eq.~\eqref{eq:coeffs_mc_schur} as a matrix and discuss their computations. 
First we can relabel elements $B^{\lm}_{q_{\lm}, q'_\lm}$ of  $B_{\rm Schur}$ as $B^{\lm}_{\Delta_\lm, q_{\lm}}$ with $\Delta_{\lm} := q'_{\lm} - q_\lm$ that can take integer values in  $\set{-2s_{\lm}, -2s_{\lm}+1, \hdots ,2 s_\lm}$. Eq.~\eqref{eq:coeffs_mc_schur} becomes 
\begin{align}
\begin{split}\label{eq:coeffs_mc_schur_reorg}
    w( B^{\lm}_{\Delta_{\lm}, q_\lm}, B^{\alpha}_{\Delta_{\alpha}, q_\alpha}) = \mathbbm{1}_{[\Delta_\lm =  -\Delta_{\alpha}]} \sum_{\substack{\mu: \\ \mu \doteq \lm + \alpha \\  q_\lm + q_\alpha \in Q_{\mu}} } \frac{\widetilde{\Pi}_{\mu, \lm, \alpha} \sqrt{d_{\lm} d_{\alpha}}}{m_{\mu}}  \left(   C_{\lm, \alpha; \mu}^{q_\lm  + \Delta_{\lm}, q_\alpha +\Delta_{\alpha} }  C_{\lm, \alpha; \mu}^{q_\lm, q_\alpha} \right)
\end{split}
\end{align}
that are non-zero only when $\Delta_\lm =  -\Delta_{\alpha}$. Hence, with appropriate ordering 
of the input and output operators  of the measurement channel, we end up with a block diagonal matrix that can be indexed by $(\Delta_\lambda, -\Delta_\lambda)$.
Finally note that $\delta$ can take values in the $\set{-n, -n+1, \hdots, n}$ such that the measurement channel in the PI Schur operator basis decomposes into $2n+1$ blocks as opposed to the $8$ blocks when expressed in the Pauli symmetrized basis.

\subsection{Constructing the classical shadow estimates}\label{app:const_cs_estimates_schur}
Akin to Sec.~\ref{app:const_cs_estimates}, to construct CS estimates of expectation values, we wish to apply the inverse measurement channel onto the operators
$(W_{\thv}^{\dag})^{\otimes n} \pihw  W_{\thv}^{\otimes n}$,
for arbitrary parameters $\thv$ and outcome $w$. To do so, we need to obtain a decomposition of these operators in $B_{\rm Schur}$. 

We recall that in vectorized form, the action by commutation of an operator $A$ onto an arbitrary operator $B$ corresponds to a matrix (superoperator) $\bar{A}$ acting on the vectorized form $\ket{B} \rangle$ of $B$. That is, $\ket{[A,B]} \rangle =  \bar{A}\ket{B} \rangle$. In turn, conjugation by the unitary $V := {\rm exp}(-it A)$ is obtained through the matrix exponentiation ${\rm exp}(-it \bar{A})$. That is, $\ket{V B V^{\dagger}} \rangle =  {\rm exp}(-it \bar{A}) \ket{B} \rangle$.

For our purposes, entries of the vector $\ket{\pihw}\rangle$ in $B_{\rm Schur}$ are readily obtained through Eq.~\eqref{eq:piw_piopsschur}. Furthermore entries of the matrices $\bar{Y}$ and $\bar{Z}$ encoding the commutation action of the generators $\widetilde{Y}=\sum_i Y_i$ and $\widetilde{Z}=\sum_i Z_i$, respectively, of any of the $W^{\otimes n}$ with $W \in \SU(2)$ were provided in Eq.~\eqref{eq:app_schur_gen_sumx_adj}.
Hence given the parametrization of Eq.~\eqref{eq:param_su2} we get 
\begin{equation}
     \ket{(W_{\thv}^{\dag})^{\otimes n} \pihw  W_{\thv}^{\otimes n}}\rangle = {\rm exp}(-i\frac{\theta_3}{2} \bar{Z}){\rm exp}(-i\frac{\theta_2}{2} \bar{Y}){\rm exp}(-i\frac{\theta_1}{2} \bar{Z}) \ket{\Pi_w}\rangle,
\end{equation}
as needed.

\clearpage
\end{document}